\documentclass[a4paper,11pt]{article}
\usepackage{jheppub}
\usepackage[dvipsnames,table]{xcolor}
\usepackage{xspace}
\usepackage[tight]{subfigure}
\usepackage{amsmath}
\usepackage{amssymb}
\usepackage{amsfonts}
\usepackage[title,titletoc]{appendix}
\usepackage{tikz}
\usepackage[normalem]{ulem}
\usepackage{physics}
\usetikzlibrary{decorations.pathmorphing}
\usetikzlibrary{decorations.markings}
\newcolumntype{C}[1]{>{\centering\arraybackslash}p{#1}}
\tikzset{
    photon/.style={decorate, decoration={snake}},
    fermion/.style={postaction={decorate},
                    decoration={markings,mark=at position .55 with {\arrow[]{>}}}},
    antifermion/.style={postaction={decorate},
                    decoration={markings,mark=at position .55 with {\arrow[]{<}}}},
    gluon/.style={decorate,
                  decoration={coil,amplitude=2pt, segment length=3pt}}
}

\def\reffi#1{\mbox{figure~\ref{#1}}}

\def\refta#1{\mbox{table~\ref{#1}}}

\def\refse#1{\mbox{section~\ref{#1}}}
\def\refapp#1{\mbox{appendix~\ref{#1}}}
\def\citere#1{\mbox{ref.~\cite{#1}}}

\newcommand{\newc}{\newcommand}
\newc{\beq}{\begin{equation}}
\newc{\eeq}{\end{equation}}
\newc{\bit}{\begin{itemize}}
\newc{\eit}{\end{itemize}}
\newc{\ben}{\begin{enumerate}}
\newc{\een}{\end{enumerate}}
\newc{\bce}{\begin{center}}
\newc{\ece}{\end{center}}
\newc{\bfi}{\begin{figure}}
\newc{\efi}{\end{figure}}

\newcommand{\ri}{\mathrm i}

\newcommand{\rT}{{\mathrm{T}}}

\newcommand{\rL}{{\mathrm{L}}}

\newcommand{\ie}{\emph{i.e.}\ }
\newcommand{\eg}{\emph{e.g.}\ }

\newcommand{\Log}{\operatorname{Log}}

\newcommand{\GeV}{\ensuremath{\,\text{GeV}}\xspace}
\newcommand{\TeV}{\ensuremath{\,\text{TeV}}\xspace}

\newcommand{\qqb}{{q\bar{q}}\xspace}
\newcommand{\PH}{\ensuremath{\text{H}}\xspace}
\newcommand{\Pj}{\ensuremath{\text{j}}\xspace}
\newcommand{\Pp}{\ensuremath{\text{p}}}
\newcommand{\Pe}{\ensuremath{\text{e}}\xspace}
\newcommand{\Pb}{\ensuremath{\text{b}}\xspace}

\newcommand{\Pt}{\ensuremath{\text{t}}\xspace}
\newcommand{\Pu}{\ensuremath{\text{u}}\xspace}
\newcommand{\Pd}{\ensuremath{\text{d}}\xspace}
\newcommand{\Ps}{\ensuremath{\text{s}}\xspace}
\newcommand{\Pc}{\ensuremath{\text{c}}\xspace}

\newcommand{\PW}{\ensuremath{\text{W}}\xspace}
\newcommand{\PZ}{\ensuremath{\text{Z}}\xspace}

\newcommand{\Mt}{\ensuremath{m_\Pt}\xspace}
\newcommand{\MH}{\ensuremath{M_\PH}\xspace}

\newcommand{\MW}{\ensuremath{M_\PW}\xspace}

\newcommand{\MVOS}{\ensuremath{M_{V}^\text{OS}}\xspace}%
\newcommand{\GVOS}{\ensuremath{\Gamma_{V}^\text{OS}}\xspace}%
\newcommand{\Mb}{\ensuremath{m_\Pb}\xspace}

\newcommand{\GH}{\ensuremath{\Gamma_\PH}\xspace}

\newcommand{\GF}{\ensuremath{G_\mu}}

\newcommand{\AvH}{{\sc AvH}\xspace}
\newcommand{\cpp}{{\sc C++}\xspace}
\newcommand{\Matrix}{{\sc Matrix}\xspace}
\newcommand{\Stripper}{{\sc Stripper}\xspace}
\newcommand{\Recola}{{\sc Recola}\xspace}
\newcommand{\vvamp}{{\sc VVamp \xspace}}
\newcommand{\OpenLoops}{O\protect\scalebox{0.8}{PEN}L\protect\scalebox{0.8}{OOPS}\xspace}

\newcolumntype{.}{D{.}{.}{-1}}
\newcolumntype{d}[1]{D{.}{.}{#1}}

\newcommand{\eqn}[1]{Eq.~(\ref{#1})}

\newcommand{\figs}[2]{Figures~\ref{#1}--\ref{#2}}

\newcommand{\as}{\alpha_{\textrm{s}}}
\newcommand{\eps}{\varepsilon}

\newcommand{\Mwo}{M^{\rm os}_{\rm{W}}}
\newcommand{\Mzo}{M^{\rm os}_{\rm{Z}}}
\newcommand{\Gzo}{\Gamma^{\rm os}_{\rm{Z}}}
\newcommand{\Gwo}{\Gamma^{\rm os}_{\rm{W}}}

\newcommand{\ptj}{p_{\rT,\Pj}}

\newcommand{\ptmiss}{p_{\rT,{\rm miss}}}

\newcommand{\ptl}{p_{\rT,\ell}}

\newcommand{\enmn}{\Pe^+\nu_\Pe\mu^-\bar\nu_\mu}
\newcommand{\ppenmn}{\Pp\Pp\to\enmn}
\dedicated{\rm CAVENDISH--HEP--21/03}
\title{NNLO QCD study of polarised ${\bf W^+W^-}$ production at the LHC}
\author{Rene Poncelet}
\author{and Andrei Popescu}

\affiliation{Cavendish Laboratory, University of Cambridge,\\
             J.J. Thomson Avenue, Cambridge CB3 0HE, United Kingdom}

\emailAdd{poncelet@hep.phy.cam.ac.uk}
\emailAdd{popescu@hep.phy.cam.ac.uk}
\date{\draftdate}

\abstract{Longitudinal polarisation of the weak bosons is a direct consequence of Electroweak
  symmetry breaking mechanism providing an insight into its nature, and is
  instrumental in searches for physics beyond the Standard Model. We perform a polarisation study of
  the diboson production in the $\ppenmn$ process at NNLO QCD in the fiducial
  setup inspired by experimental measurements at ATLAS. This is the first
  polarisation study at NNLO. We employ the double-pole approximation framework
  for the polarised calculation, and investigate NNLO effects arising in
  differential distributions.}

\keywords{Electroweak bosons, Polarisation, NNLO QCD, Diboson, LHC}

\begin{document}

\maketitle

\section{Introduction}\label{sec:introduction}

Weak boson polarisation is under intense research both on the theoretical and
experimental side. It is a handle to directly probe the Standard Model (SM)
electroweak (EW) symmetry breaking mechanism and is instrumental in constraining
the triple and quartic gauge boson couplings for beyond SM physics searches.

Several processes have been studied theoretically in the context of weak boson polarisation.
Seminal papers covered the $V$+j process \cite{Bern:2011ie,Stirling:2012zt}. Later on
other processes were considered, such as diboson production
\cite{Baglio:2018rcu,Baglio:2019nmc,Denner:2020bcz,Denner:2020eck}, vector boson
scattering (VBS) \cite{Ballestrero:2017bxn,Ballestrero:2019qoy,De:2020iwq}.
Top-quark decays are currently under investigation.

The amount of statistics of Run 2 at the LHC has enabled polarised measurement
in relatively high cross section processes such as $V$+jet
\cite{Chatrchyan:2011ig,ATLAS:2012au,Khachatryan:2015paa,Aad:2016izn}, boson
\cite{Aaboud:2019nkz,Aaboud:2019gxl} and top-quark pair production
\cite{Aaboud:2016hsq,Khachatryan:2016fky}. There are good prospects for
polarised VBS signals at high-luminosity LHC \cite{Azzi:2019yne} and there
already are some results available \cite{Sirunyan:2020gvn}. It is impossible to
directly select bosons with a specified polarisation, but methods like
reweighting procedures
\cite{Chatrchyan:2011ig,ATLAS:2012au,Khachatryan:2015paa,Aad:2013ksa} and
artificial intelligence techniques
\cite{Searcy:2015apa,Lee:2018xtt,Lee:2019nhm,Grossi:2020orx} allow for analysis
of polarised signals. The main result is the extracted polarisation
coefficients which are then compared to theoretical values. Close attention is
paid to differential distributions for longitudinally polarised bosons, which
is a direct probe of the EW symmetry breaking mechanism.

The production of weak boson pairs has been extensively studied in the literature.
Next-to-leading order (NLO)
\cite{Billoni:2013aba,Biedermann:2016guo,Bierweiler:2012kw},
next-to-next-to-leading order (NNLO)
\cite{Caola:2015rqy,Grazzini:2016ctr,Kallweit:2019zez} and combined NLO EW and
(N)NLO QCD \cite{Denner:2019tmn,Denner:2020zit} computations are available for
a variety of setups and observables. Resummation and parton shower effects
have also been studied in the context of weak boson pair production
\cite{Re:2018vac,Kallweit:2020gva,Brauer:2020kfv,Chiesa:2020ttl,Kuhn:2011mh}.
Recent progress has been made in the computation of NLO corrections to cross
sections for polarised bosons \cite{Denner:2020bcz,Denner:2020eck}.  There are
two main obstacles that are in the way of direct theoretical calculations with
polarised boson in the experimentally accessible signatures. Firstly, weak
bosons are short-lived particles, and they can be observed only through their
leptonic and hadronic decay signatures. They are produced off-shell and some
adjustment is required to make sense of their polarisation state. Secondly, the
signatures involve a non-resonant\footnote{Non-resonant as opposed to
double-resonant. More precisely, single-resonant.}
background which cannot be removed in a
simple manner, because it is essential for gauge invariance of the whole
amplitude. Effects of the gauge invariance breakdown are severe
\cite{Ballestrero:2019qoy}.  The commonly used approach to tackle both issues
is to use on-shell amplitudes which can be obtained either by restricting the
integration phase space in the way of the narrow-width Approximation
\cite{Denner:1999gp,Denner:2005fg}
or by means of an on-shell projection (OSP)
\cite{Ballestrero:2017bxn,Ballestrero:2019qoy}, also known as pole
approximation. Regardless of the particular implementation, a method that uses
on-shell amplitudes has an intrinsic uncertainty of
$\mathcal{O}(\frac{\Gamma}{M})$ but it is still advantageous in comparison with
the indirect approach involving reweighting which is used in experimental
analysis as shown in the case of $\PW\PZ$ production \cite{Ballestrero:2019qoy}.

In this paper we address, for the first time, NNLO QCD correction for the polarised
$\PW^+\PW^-$ production. We compute fiducial and differential cross sections at NNLO
QCD accuracy for the LHC at 13 \TeV and investigate what are the effects at
this precision level. NNLO corrections are particularly important for the
differential distributions in diboson production, where NLO scale uncertainty
exceeds the intrinsic uncertainty related to the theoretical definition of
boson polarisation. Additionally, we explore how narrow-width approximation
performs in comparison with double-pole approximation which is assumed to be
more accurate due to its incorporation of off-shell effects.

The structure of this paper is as follows. In
\refse{subsec:polarised_weak_bosons} we discuss approaches which we apply to
define boson polarisations in the diboson production process. Then we specify
our setup including the SM parameters, selection cuts, and a list of
computational tools that we use. \refse{sec:results} is dedicated to our
results. We present the integrated cross sections, and discuss the pure NNLO QCD
corrections in \refse{subsec:pure_nnlo_effects}. We add the loop-induced
channel and discuss its effects in
\refse{subsec:effects_of_loop_induced_contribution}.  In
\refse{subsec:comparison_between_dpa_and_nwa} we compare the
narrow-width approximation and
the double-pole approximation for unpolarised weak bosons against an off-shell
computation.  In \refse{sec:conclusion} we summarise our findings.

\section{Details of the calculation}
\subsection{Polarised weak bosons}\label{subsec:polarised_weak_bosons}

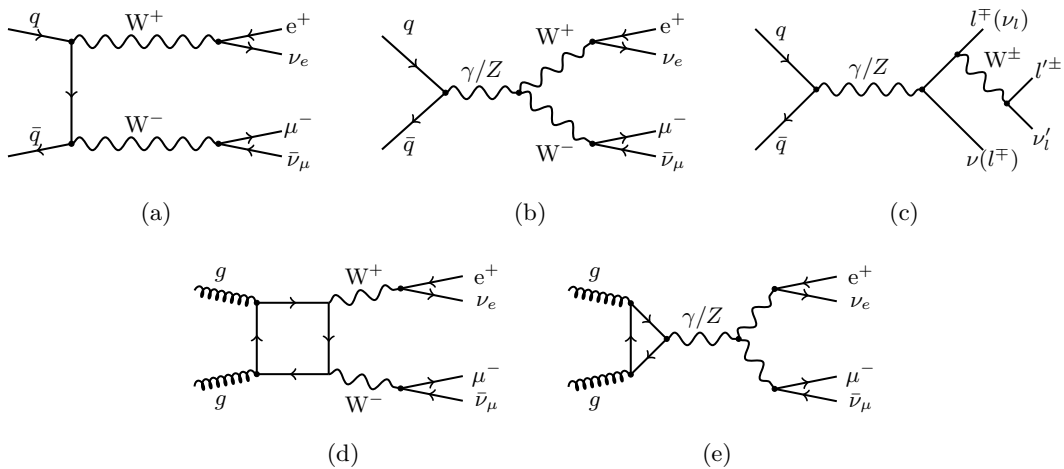
\begin{figure}[!ht]
  \centering
  \subfigure[]{
    \resizebox{0.3\textwidth}{!}{\begin{tikzpicture}[scale=1]
  \def \inX    {-2}
  \def \inDY   {.2}
  \def \bosonX {-1}
  \def \bosonY {.7}
  \def \decayX {1}
  \def \lepDX  {1}
  \def \lepDY  {.2}
  \def \vsize  {1pt}
  \def \Wp {\bosonX, \bosonY}
  \def \Wm {\bosonX,-\bosonY}
  \def \Wpd {\decayX, \bosonY}
  \def \Wmd {\decayX,-\bosonY}
  \node (p1) at  (\inX, \bosonY+\inDY) {};
  \node (p2) at  (\inX, -\bosonY-\inDY) {};
  \node (p3) at  (\decayX+\lepDX, \bosonY+\lepDY) {};
  \node (p4) at  (\decayX+\lepDX, \bosonY-\lepDY) {};
  \node (p5) at  (\decayX+\lepDX,-\bosonY+\lepDY) {};
  \node (p6) at  (\decayX+\lepDX,-\bosonY-\lepDY) {};
  \filldraw (\Wp) circle (\vsize);
  \filldraw (\Wm) circle (\vsize);
  \filldraw (\Wpd) circle (\vsize);
  \filldraw (\Wmd) circle (\vsize);
  \node at ({.5*(\inX+\bosonX)}, \bosonY+1.5*\inDY) {\footnotesize $q$};
  \node at ({.5*(\inX+\bosonX)}, -\bosonY+.5*\inDY) {\footnotesize $\bar q$};
  \node at ({.5*(\decayX+\bosonX)}, \bosonY+.3) {\footnotesize $\PW^+$};
  \node at ({.5*(\decayX+\bosonX)}, -\bosonY+.3) {\footnotesize $\PW^-$};
  \node at (\decayX+\lepDX+.1, \bosonY+\lepDY+.05) {\footnotesize $\Pe^+$};
  \node at (\decayX+\lepDX+.1, \bosonY-\lepDY-.05) {\footnotesize $\nu_e$};
  \node at (\decayX+\lepDX+.1, -\bosonY+\lepDY+.05) {\footnotesize $\mu^-$};
  \node at (\decayX+\lepDX+.1, -\bosonY-\lepDY-.05) {\footnotesize $\bar\nu_\mu$};
  \draw [thick,fermion]     (p1) to (\Wp) {};
  \draw [thick,antifermion] (p2) to (\Wm) {};
  \draw [thick,fermion]     (\Wp) to (\Wm) {};
  \draw [thick,photon]  (\Wp) to (\Wpd) {};
  \draw [thick,photon]  (\Wm) to (\Wmd) {};
  \draw [thick,antifermion] (\Wpd) to (p3) {};
  \draw [thick,fermion]     (\Wpd) to (p4) {};
  \draw [thick,fermion]     (\Wmd) to (p5) {};
  \draw [thick,antifermion] (\Wmd) to (p6) {};
\end{tikzpicture}}
    \label{subfig:dr-ab}
  }
  \subfigure[]{
    \resizebox{0.3\textwidth}{!}{\begin{tikzpicture}[scale=1]
  \def \inX    {-2}
  \def \inDY   {.2}
  \def \bosonX {-1}
  \def \bosonXb {0}
  \def \bosonY {.7}
  \def \decayX {1}
  \def \lepDX  {1}
  \def \lepDY  {.2}
  \def \vsize  {1pt}
  \def \Zprod  {\bosonX, 0}
  \def \Zdecay {\bosonXb,0}
  \def \Wpd {\decayX, \bosonY}
  \def \Wmd {\decayX,-\bosonY}
  \node (p1) at  (\inX, \bosonY+\inDY) {};
  \node (p2) at  (\inX, -\bosonY-\inDY) {};
  \node (p3) at  (\decayX+\lepDX, \bosonY+\lepDY) {};
  \node (p4) at  (\decayX+\lepDX, \bosonY-\lepDY) {};
  \node (p5) at  (\decayX+\lepDX,-\bosonY+\lepDY) {};
  \node (p6) at  (\decayX+\lepDX,-\bosonY-\lepDY) {};
  \filldraw (\Zprod) circle (\vsize);
  \filldraw (\Zdecay) circle (\vsize);
  \filldraw (\Wpd) circle (\vsize);
  \filldraw (\Wmd) circle (\vsize);
  \node at ({.5*(\inX+\bosonX)}, \bosonY+1.0*\inDY) {\footnotesize $q$};
  \node at ({.5*(\inX+\bosonX)}, -\bosonY-.2*\inDY) {\footnotesize $\bar q$};
  \node at ({.5*(\bosonX+\bosonXb)},.3) {\footnotesize $\gamma/Z$};
  \node at ({.5*(\bosonXb+\decayX)}, \bosonY+0.1) {\footnotesize $\PW^+$};
  \node at ({.5*(\bosonXb+\decayX)}, -\bosonY-0.1) {\footnotesize $\PW^-$};
  \node at (\decayX+\lepDX+.1, \bosonY+\lepDY+.05) {\footnotesize $\Pe^+$};
  \node at (\decayX+\lepDX+.1, \bosonY-\lepDY-.05) {\footnotesize $\nu_e$};
  \node at (\decayX+\lepDX+.1, -\bosonY+\lepDY+.05) {\footnotesize $\mu^-$};
  \node at (\decayX+\lepDX+.1, -\bosonY-\lepDY-.05) {\footnotesize $\bar\nu_\mu$};
  \draw [fermion,thick]     (p1) to (\Zprod) {};
  \draw [antifermion,thick] (p2) to (\Zprod) {};
  \draw [thick,photon]      (\Zprod) to (\Zdecay) {};
  \draw [thick,photon]      (\Zdecay) to (\Wpd) {};
  \draw [thick,photon]      (\Zdecay) to (\Wmd) {};
  \draw [antifermion,thick] (\Wpd) to (p3) {};
  \draw [fermion,thick]     (\Wpd) to (p4) {};
  \draw [fermion,thick]     (\Wmd) to (p5) {};
  \draw [antifermion,thick] (\Wmd) to (p6) {};
\end{tikzpicture}}
    \label{subfig:dr-f}
  }
  \subfigure[]{
    \resizebox{0.3\textwidth}{!}{  \begin{tikzpicture}[scale=1]
  \def \inX      {-2.0}
  \def \inY      { 1.0}
  \def \ZprodX   {-1.0}
  \def \ZdecayX  {  .5}
  \def \WprodX   { 1.0}
  \def \WprodY   { 0.5}
  \def \WdecayX  { 1.7}
  \def \WdecayY  {-0.2}
  \def \vsize    {1pt}
  \def \lepDX    { .5}
  \def \lepDY    { .5}
  \def \Zprod  {\ZprodX,         0}
  \def \Zdecay {\ZdecayX,        0}
  \def \Wprod  {\WprodX,   \WprodY}
  \def \Wdecay {\WdecayX, \WdecayY}
  \node (p1) at  (\inX,  \inY) {};
  \node (p2) at  (\inX, -\inY) {};
  \node (p3) at  ({2*(\WprodX)-\ZdecayX}, 2*\WprodY) {};
  \node (p4) at  ({2*(\WprodX)-\ZdecayX}, -2*\WprodY) {};
  \node (p5) at  (\WdecayX+\lepDX,\WdecayY+\lepDY) {};
  \node (p6) at  (\WdecayX+\lepDX,\WdecayY-\lepDY) {};
  \filldraw (\Zprod) circle (\vsize);
  \filldraw (\Zdecay) circle (\vsize);
  \filldraw (\Wprod) circle (\vsize);
  \filldraw (\Wdecay) circle (\vsize);
  \node at ({.5*(\inX+\ZprodX)}, { 0.5*(\inY)+.3})  {\footnotesize $q$};
  \node at ({.5*(\inX+\ZprodX)}, {-0.5*(\inY)-.3}) {\footnotesize $\bar q$};
  \node at ({2*(\WprodX)-\ZdecayX+.1},  2*\WprodY) {\footnotesize $l^{\mp}(\nu_l)$};
  \node at ({2*(\WprodX)-\ZdecayX}, -2*\WprodY) {\footnotesize $\nu(l^{\mp})$};
  \node at ({\WdecayX+\lepDX+.1},  \WdecayY+\lepDY) {\footnotesize $l'^{\pm}$};
  \node at ({\WdecayX+\lepDX},  \WdecayY-\lepDY) {\footnotesize $\nu'_l$};
  \node at ({.5*(\ZprodX+\ZdecayX)},.3) {\footnotesize $\gamma/Z$};
  \node at ({.5*(\WprodX+\WdecayX)+.3}, {.5*(\WprodY+\WdecayY)+.3}) {\footnotesize $\PW^\pm$};
  \draw [fermion,thick]     (p1) to (\Zprod) {};
  \draw [antifermion,thick] (p2) to (\Zprod) {};
  \draw [thick,photon]      (\Zprod) to (\Zdecay) {};
  \draw [thick,photon]      (\Wprod) to (\Wdecay) {};
  \draw [thick] (\Zdecay) to (\Wprod) {};
  \draw [thick] (\Wprod) to (p3) {};
  \draw [thick]     (\Zdecay) to (p4) {};
  \draw [thick]     (\Wdecay) to (p5) {};
  \draw [thick] (\Wdecay) to (p6) {};
\end{tikzpicture}}
    \label{subfig:sr-general}
  }
    \\
    \subfigure[]{
    \resizebox{0.3\textwidth}{!}{  \begin{tikzpicture}[scale=1]
  \def \inX    {-2}
  \def \inDY   { 0.2}
  \def \qX     {-1.0}
  \def \qY     { 0.5}
  \def \WprodX { 0.0}
  \def \WprodY { \qY}
  \def \WdecayX{ 1.0}
  \def \WdecayY{ 0.7}
  \def \lepDX  { 1}
  \def \lepDY  { 0.2}
  \def \vsize  {1pt}
  \def \Wp {\WprodX, \qY}
  \def \Wm {\WprodX,-\qY}
  \def \Wpd {\WdecayX, \WdecayY}
  \def \Wmd {\WdecayX, -\WdecayY}
  \node (p1) at  (\inX, \qY+\inDY) {};
  \node (p2) at  (\inX, -\qY-\inDY) {};
  \node (p3) at  (\WdecayX+\lepDX, \WdecayY+\lepDY) {};
  \node (p4) at  (\WdecayX+\lepDX, \WdecayY-\lepDY) {};
  \node (p5) at  (\WdecayX+\lepDX,-\WdecayY+\lepDY) {};
  \node (p6) at  (\WdecayX+\lepDX,-\WdecayY-\lepDY) {};
  \filldraw (\qX, \qY) circle (\vsize);
  \filldraw (\qX,-\qY) circle (\vsize);
  \filldraw (\Wpd) circle (\vsize);
  \filldraw (\Wmd) circle (\vsize);
  \draw [thick,gluon]     (p1) to (\qX, \qY) {};
  \draw [thick,gluon]     (p2) to (\qX,-\qY) {};
  \draw [thick,fermion]     (\qX,\qY) to (\Wp) {};
  \draw [thick,fermion]     (\Wp) to (\Wm) {};
  \draw [thick,antifermion] (\qX,-\qY) to (\Wm) {};
  \draw [thick,fermion]     (\qX,-\qY) to (\qX,\qY) {};
  \draw [thick,photon]  (\Wp) to (\Wpd) {};
  \draw [thick,photon]  (\Wm) to (\Wmd) {};
  \draw [thick,antifermion] (\Wpd) to (p3) {};
  \draw [thick,fermion]     (\Wpd) to (p4) {};
  \draw [thick,fermion]     (\Wmd) to (p5) {};
  \draw [thick,antifermion] (\Wmd) to (p6) {};
  \node at ({.5*(\inX+\qX)}, \qY+.4) {\footnotesize $g$};
  \node at ({.5*(\inX+\qX)},-\qY-.4) {\footnotesize $g$};
  \node at ({(\WprodX+\WdecayX)*.5}, {(\WprodY+\WdecayY)*.5+.3}) {\footnotesize $\PW^+$};
  \node at ({(\WprodX+\WdecayX)*.5}, {-(\WprodY+\WdecayY)*.5-.3}) {\footnotesize $\PW^-$};
  \node at (\WdecayX+\lepDX+.2, \WdecayY+\lepDY) {\footnotesize $\Pe^+$};
  \node at (\WdecayX+\lepDX+.2, \WdecayY-\lepDY) {\footnotesize $\nu_e$};
  \node at (\WdecayX+\lepDX+.2,-\WdecayY+\lepDY) {\footnotesize $\mu^-$};
  \node at (\WdecayX+\lepDX+.2,-\WdecayY-\lepDY) {\footnotesize $\bar\nu_\mu$};
\end{tikzpicture}}
    \label{subfig:li-a}
  }
  \subfigure[]{
    \resizebox{0.3\textwidth}{!}{\begin{tikzpicture}[scale=1]
  \def \inX    {-2}
  \def \inDY   { 0.2}
  \def \qX     {-1}
  \def \qY     { 0.5}
  \def \ZprodX {-0.5}
  \def \ZdecayX{ 0.5}
  \def \WdecayX{ 1.0}
  \def \WdecayY{ 0.7}
  \def \lepDX  { 1}
  \def \lepDY  { 0.2}
  \def \vsize  {1pt}
  \def \Zprod {\ZprodX, 0}
  \def \Zdecay {\ZdecayX, 0}
  \def \Wpd {\WdecayX, \WdecayY}
  \def \Wmd {\WdecayX, -\WdecayY}
  \node (p1) at  (\inX, \qY+\inDY) {};
  \node (p2) at  (\inX, -\qY-\inDY) {};
  \node (p3) at  (\WdecayX+\lepDX, \WdecayY+\lepDY) {};
  \node (p4) at  (\WdecayX+\lepDX, \WdecayY-\lepDY) {};
  \node (p5) at  (\WdecayX+\lepDX,-\WdecayY+\lepDY) {};
  \node (p6) at  (\WdecayX+\lepDX,-\WdecayY-\lepDY) {};
  \filldraw (\qX, \qY) circle (\vsize);
  \filldraw (\qX,-\qY) circle (\vsize);
  \filldraw (\Zprod) circle (\vsize);
  \filldraw (\Zdecay) circle (\vsize);
  \filldraw (\Wpd) circle (\vsize);
  \filldraw (\Wmd) circle (\vsize);
  \draw [thick,gluon]     (p1) to (\qX, \qY) {};
  \draw [thick,gluon]     (p2) to (\qX,-\qY) {};
  \draw [thick,fermion]     (\qX,\qY) to (\Zprod) {};
  \draw [thick,fermion]     (\Zprod) to (\qX,-\qY) {};
  \draw [thick,fermion]     (\qX,-\qY) to (\qX,\qY) {};
  \draw [thick,photon]  (\Zprod) to (\Zdecay) {};
  \draw [thick,photon]  (\Zdecay) to (\Wpd) {};
  \draw [thick,photon]  (\Zdecay) to (\Wmd) {};
  \draw [thick,antifermion] (\Wpd) to (p3) {};
  \draw [thick,fermion]     (\Wpd) to (p4) {};
  \draw [thick,fermion]     (\Wmd) to (p5) {};
  \draw [thick,antifermion] (\Wmd) to (p6) {};
  \node at ({.5*(\inX+\qX)}, \qY+.4) {\footnotesize $g$};
  \node at ({.5*(\inX+\qX)},-\qY-.4) {\footnotesize $g$};
  \node at ({.5*(\ZprodX+\ZdecayX)},.3) {\footnotesize $\gamma/Z$};
  \node at (\WdecayX+\lepDX+.2, \WdecayY+\lepDY) {\footnotesize $\Pe^+$};
  \node at (\WdecayX+\lepDX+.2, \WdecayY-\lepDY) {\footnotesize $\nu_e$};
  \node at (\WdecayX+\lepDX+.2,-\WdecayY+\lepDY) {\footnotesize $\mu^-$};
  \node at (\WdecayX+\lepDX+.2,-\WdecayY-\lepDY) {\footnotesize $\bar\nu_\mu$};
\end{tikzpicture}}
    \label{subfig:li-f}
  }
  \caption{Selected diagrams contributing to the $\ppenmn$ process.  Diagrams
    (a,b) represent Born double-resonant contribution; (c) -- general case of
    Born single-resonant contributions (background); (d,e) -- loop-induced
    double-resonant contribution;
    \label{fig:process-diagrams}}
\end{figure}

In this paper we study the resonant production and subsequent decay of
(un)polarised $W^+W^-$-boson pairs at the LHC in the different flavour
di-leptonic decay channel, i.e.
\begin{equation}
 \Pp\Pp \to \PW^+\PW^- + X \to \enmn + X \;.
\end{equation}
\reffi{subfig:dr-ab} and \reffi{subfig:dr-f} show the contributing
\emph{double-resonant} Feynman diagrams.
The resonant process is a part of the more general off-shell production of the same
leptonic final state
\begin{equation}
  \ppenmn + X.
\end{equation}
This process has additional contributions from \emph{single-resonant} diagrams,
see \reffi{subfig:sr-general}. To define the production and decay of
\emph{polarised} intermediate vector-bosons in a gauge-invariant way both
$\PW$-bosons are required to be on their mass-shell. We consider two
commonly used approximations both resulting in on-shell amplitudes for
polarised $\PW$-bosons: the so-called pole approximation or, in this case,
\emph{double-pole approximation} (DPA), and the \emph{narrow-width approximation}
(NWA). Both methods neglect \emph{single-resonant} contributions present in the
general process $\ppenmn$ and introduce uncertainties which are formally of
$\order{\Gamma_\PW/M_\PW}$. While this error estimate holds for inclusive
observables, the uncertainty in differential distributions can be significantly
larger.

NWA considers the limit $\Gamma_\PW/m_\PW \to 0$ in the cross section, thus
neglecting $\mathcal{O}(\Gamma_\PW/m_\PW)$ terms, rendering the $\PW$-bosons
on-shell, and factorizing the amplitudes and phase spaces of production and
decay. The NWA works well with massive short-lived particles such as
weak bosons and top-quarks, and so it is well suited for this study. By
construction, this approximation performs poorly for observables which are
sensitive to the off-shellness of the vector bosons. There exist
extensions to the NWA, which attempt to include off-shell effects, such as the
Madspin approach \cite{BuarqueFranzosi:2019boy}, which simulates the off-shell
effects by using the Breit-Wigner sampling for the resonant propagator.  We do
not consider such an extension here. The production and decay are correlated
through the boson's momenta and polarisations, schematically the amplitude
factorizes as follows:
\begin{equation}\label{eq:NWA}
 \mathcal{M}_{{\ppenmn}} \sim \sum_{h,h' \in \Lambda}
   \mathcal{M}^{h,h'}_{{\Pp\Pp\to \PW^+\PW^-}}\Gamma^h_{{\PW^+\to \Pe^+\nu_\Pe}}
       \Gamma^{h'}_{{\PW^-\to \mu^-\bar{\nu}_{\mu}}},
\end{equation}
where $h,h' \in \Lambda = \{+,-,L\}$ stand for the $\PW$-boson polarisations.
By restricting the sum to specific choices for $h,h'$ we define the polarised
production and decay. We denote the coherent sum of the transverse
polarisations $\{+,-\}$ as (T) and the longitudinal polarisation as (L).

The DPA \cite{Beenakker:1998gr,Billoni:2013aba,Denner:2019vbn} approach instead considers the
off-shell phase space and introduces an approximation for the amplitudes alone.
In order to guarantee gauge invariance, one defines an on-shell projection to map
the off-shell kinematics of the decay products point-by-point to the on-shell
kinematics. This allows the same factorization as in \eqn{eq:NWA} by neglecting
\emph{single-resonant} diagrams. Boson propagators with the off-shell
kinematics are kept for modelling the Breit-Wigner shape of the off-shell
amplitude.

DPA has an advantage over NWA in that it generates the off-shell kinematics,
however it is not uniquely defined. The on-shell mapping has to be specified,
whereas NWA approach is unambiguous. We will compare performance of these two
methods in \refse{subsec:comparison_between_dpa_and_nwa}.

For the polarisation study at NNLO we will use the DPA approach. We follow
\cite{Ballestrero:2017bxn} in defining the on-shell projection, where the
following conserved quantities are suggested:
\begin{enumerate}
  \item the total diboson momentum $p_{\PW^+\PW^-}$;
  \item the direction of $p_{\PW^+}$ momentum in diboson centre-of-mass frame
    (after a direct boost in Lab frame);
  \item the angles of charged leptons w.r.t to their parent boson momentum (in Lab
    frame) in their parent $\PW$-boson centre-of-mass frame (after a direct boost
    in Lab frame).
\end{enumerate}

The algorithm goes as follows.
Consider the diboson mass frame by a direct boost from Lab frame.
In this frame individual boson momenta are equal and back-to-back, but
generally not on-shell. To correct this, for each boson momenta, we fix the
energy to be $\sqrt{s}/2$ and rescale the spatial part so that its length
becomes $\frac{1}{2} \sqrt{s - 4M_\PW^2}$, while the angles are kept untouched.
These modifications do not affect the total momentum of the diboson system.
Next, we turn to the decay products. In order to modify momenta of ${\Pe^+},
{\nu_\Pe}$ we reconstruct helicity frames in $\PW^+$-boson rest frame, and
calculate angles of the positron using the original off-shell kinematics.
While polar angle definition is unambiguous, the rotation of the XY-plane and
thus the azimuthal angle is subject to specification. We discuss the azimuthal
angle definition we use in \refapp{sec:appendix_azimuth}. With the new on-shell
$\PW^+$-boson momentum we construct the new positron momentum in the new parent
boson rest frame with the original polar and azimuthal angles. The neutrino
momentum is trivially inferred. Analogously, we build new momenta for the decay
products of $\PW^-$.

The process under consideration allows for unambiguous on-shell mapping,
but if there exists an ambiguity around combining the decay products
into parent resonances, such as appearing in $\PZ\PZ$ production or NLO EW radiation,
the OSP should be revised. A treatment of non-factorisable corrections and a generic massive particle
configuration can be found in \cite{Dittmaier:2015bfe}.

The definition of the boson polarisation vectors entering the polarised production
and decay amplitudes is not unambiguous and needs to be chosen. A particular choice,
which we employ in this work, based on momenta in the laboratory frame, is
\begin{align}
\eps^{\mu}_{-} &= \frac{1}{\sqrt 2}
  \left(0,\cos\theta_V\cos\phi_V +  \ri\sin\phi_V , \cos\theta_V\sin\phi_V -
  \ri\cos\phi_V, - \sin\theta_V \right) \,,\notag \\
\eps^{\mu}_{+} &= \frac{1}{\sqrt 2}
    \left(0,- \cos\theta_V\cos\phi_V +  \ri\sin\phi_V ,
       - \cos\theta_V\sin\phi_V - \ri\cos\phi_V, \sin\theta_V\right) \,,\notag \\
\eps^{\mu}_{\rL} &=
    \frac{1}{M}\left(p,E \sin\theta_V\cos\phi_V,E \sin\theta_V\sin\phi_V,E
    \cos\theta_V\right),
\end{align}
for left, right, and longitudinal polarisations respectively. Here $M, p, E$ are
mass, total momentum, and energy of the weak boson, and $\theta_V, \phi_V$ are
its angles in a selected frame.

In this study we define polarisation vectors in the laboratory frame which is
more accessible experimentally. However there exist other alternatives, \eg
the diboson centre-of-mass frame, which was also used in the experimental
studies \cite{Sirunyan:2020gvn}. It has been observed that the frame choices
tend to be rather complementary to each other in their discrimination power to
isolate polarisations \cite{Denner:2020eck}.

\subsection{Numerical parameters}\label{subsec:numerical_parameters}

To fully specify our computational setup we give a summary of all numerical
input parameters. We use the following set of particle parameters:
\begin{alignat}{2}\label{smparqq}
 \Mwo &= 80.3790 \GeV,&\qquad \Gwo &= 2.0850\GeV, \notag\\
 \Mzo &= 91.1876 \GeV,&\qquad \Gzo &= 2.4952\GeV, \notag\\
 \MH &= 125 \GeV,&\qquad \GH &= 0.00407\GeV, \notag\\
 \Mt &= 173 \GeV,&\quad \Mb &= 4.7\GeV;
\end{alignat}
where the boson parameters actually used in the calculation (pole values) are
obtained by means of \cite{Bardin:1988xt},
\beq M_V =
  \frac{\MVOS}{\sqrt{1+(\GVOS/\MVOS)^2}}\,,\qquad \Gamma_V =
  \frac{\GVOS}{\sqrt{1+(\GVOS/\MVOS)^2}};
\eeq
for $V = \PW, \PZ$. All leptons are considered massless, which makes the results
insensitive to the specific lepton flavours as long as they belong to different
generations in order for the diboson reconstruction to remain unique. All
other quarks ($\Pu,\Pd,\Pc,\Ps$) are considered massless.

We consider the 5-flavour PDF set \texttt{NNPDF31\_[n]nlo\_as\_0118} (IDs:
303400, 303600) approximation for [N]NLO \cite{Tanabashi:2018oca} as
implemented in LHAPDF \cite{Buckley:2014ana}.  However, we use massive bottom
quarks throughout the calculation to avoid contributions from off-shell
top-quark pair production which would enter at NNLO and would be regarded as a
separate process. Real emission contribution of massive $b\bar{b}$ quarks
are neglected for the same reason. The numerical impact of the mismatch in the number of
massless quarks between the PDF and the perturbative part
is up to 0.6\% at the total cross section level and up to 8\% in the distribution tails
(\eg lepton $p_T$), however it is within by the factorisation scale uncertainty band, as
we estimated at NLO.
We would also like to point out that the scheme we use is~
--- up to NLO --- effectively the same as removing processes with a $b$-quark in
the initial state, so with the chosen PDF set we are able to directly compare our
results with \cite{Denner:2020bcz} at NLO.

We use the complex mass scheme framework \cite{Denner:2006ic} and the
couplings are fixed following the $G_\mu$ scheme with
\begin{equation}
 \GF = 1.16638\cdot10^{-5} \GeV^{-2}.
\end{equation}
Within the frameworks of NWA and DPA we set weak boson ($\PW$ \emph{and} $\PZ$) widths to zero
in the calculation of couplings and the Weinberg angle, so they remain real.

Both factorisation and renormalisation scales are set to $\PW$ pole mass:
$\mu_F = \mu_R = \MW$.

The cuts we use are presented as fiducial setup in \citere{Denner:2020bcz}
inspired by ATLAS measurements \cite{Aaboud:2019nkz}:
\begin{itemize}
\item minimum transverse momentum of the charged leptons, $\ptl>27\GeV$;
\item maximum rapidity of the charged leptons, $|y_\ell|<2.5$;
\item minimum missing transverse momentum, $\ptmiss> 20\GeV$;
\item veto on events containing at least one jet candidate
  with $\ptj>35\GeV,|\eta_{\Pj}|<4.5$;
\item minimum invariant mass of the charged lepton-pair system,
  $M_{\Pe^+\mu^-}>55\GeV$.
\end{itemize}
Also, by construction, DPA and NWA contain an implicit cut on diboson invariant
mass: $M_{\PW^+\PW^-} > 2\cdot M_\PW$.

The jet veto is used to reduce giant K-factors \cite{Kallweit:2019zez} otherwise
mostly appearing at NLO but also driving up NNLO corrections. CKM matrix is
assumed to be diagonal. Finally, the invariant mass cut reduces the Higgs
background in the $gg$-initiated process. Note that we did not apply the
$M_{\PW^+\PW^-} > 130 \GeV$ cut for the $gg$-initiated process as suggested in
\cite{Denner:2020bcz} to exclude the Higgs peak region as it has little effect
on the results.

\subsection{Tools used in the calculation}\label{subsec:tools_used}

The computation has been done using \Stripper framework, a \cpp implementation
of the four-dimensional formulation of the sector-improved residue subtraction
scheme \cite{Czakon:2010td,Czakon:2014oma}. \Stripper is a library which supplies a
Monte-Carlo generator and automates the subtraction scheme, and it relies on
external tools for calculating tree-level, one-loop and two-loop amplitudes. It
has been successfully applied to the production of top-quark pairs
\cite{Behring:2019iiv,Czakon:2020qbd}, inclusive jets \cite{Czakon:2019tmo},
and three photons \cite{Chawdhry:2019bji}. We use \AvH library
\cite{Bury:2015dla} to provide the
Born amplitudes. The one-loop amplitudes are calculated using \OpenLoops 2
\cite{Buccioni:2019sur,Cascioli:2011va,Buccioni:2017yxi}, which we modified to
extend its functionality to specify weak boson polarisations.  The two-loop
amplitudes for $\qqb$-induced channel were provided by \vvamp project
\cite{Gehrmann:2015}.

Several checks have been performed both on the integrated cross section level and
per phase space point. The total cross section for the off-shell setup
calculated within Stripper framework was checked against \Matrix at NNLO
\cite{Gehrmann:2014fva} in the inclusive setup. Our private build of \OpenLoops
2 \cite{Buccioni:2019sur} was checked on the amplitude level against the private
version of \Recola used in \cite{Denner:2020bcz} for various DPA setups. We
checked our differential distributions at NLO against the ones provided in
\cite{Denner:2020bcz} as well as the total cross section results for various
polarised setups.

\section{Results}\label{sec:results}

In this section, we present phenomenological results for the polarised
signals in $\PW$-pair production in the fiducial setup on the LHC at a
hadronic centre of mass energy of $13\TeV$.

Diboson production and its further decay into leptons is represented by the
diagrams in \reffi{fig:process-diagrams}. The loop-induced contribution in
\figs{subfig:li-a}{subfig:li-f} enters the calculation for the first time at
NNLO in $\as$. It is effectively a LO contribution which introduces substantial
corrections. In what follows we will refer to NNLO corrections to diagrams in
\figs{subfig:dr-ab}{subfig:sr-general} as to 'NNLO (without LI)', or just
'NNLO', and to corrections that include the loop-induced channel as to 'NNLO
(with LI)' or 'NNLO+LI' corrections. The polarisation setups are identified by
its polarised boson.  We will abbreviate polarisation setups by two letters out
of the set \{U, T, L\}, which correspond to unpolarised, transverse, and
longitudinal boson polarisation respectively. For example, the singly-polarised
setup with longitudinal $\PW^+$ boson will read 'LU'.

We provide LO, NLO, and NNLO results, the NNLO K-factor
which is calculated as $\sigma_{\text{NNLO}}/\sigma_{\text{NLO}}$ at the central scale, and the
NNLO+LI which includes the loop-induced channel. Scale uncertainties are calculated using
the standard independent 7-point variation of $\mu_R, \mu_F$ by a factor of $2$
around the central scale, with the restriction $1/2 \leq \mu_R/\mu_F \leq 2$.

After discussing the fiducial cross section in the next section, we will turn
our focus on 'NNLO (without LI)' corrections to differential distributions
\refse{subsec:pure_nnlo_effects}. The effects of the loop-induced channel on
differential distributions will be explored in
\refse{subsec:effects_of_loop_induced_contribution}. Finally, a comparison on
differential level of the DPA and NWA approaches against the off-shell
computation will be performed in \refse{subsec:comparison_between_dpa_and_nwa}.

\subsection{Fiducial cross sections}\label{subsec:fiducial_cross_sections}

The total cross-section results for various polarisation setups are presented
in \refta{table:totalxsec}. It also includes unpolarised calculations performed
in the frameworks of DPA, NWA, and the off-shell calculation. Scale
uncertainties are presented in percentage values with respect to the central
scale result as sub- and superscripts. Monte-Carlo numerical errors on the
central scale values are indicated in parentheses and correspond to the last
significant digit of the result.

In the unpolarised setups we see that DPA undershoots the off-shell calculation
by 2.5\%. DPA is not supposed to fully match the off-shell calculation as it
only includes double-resonant contributions relevant for the diboson
production. This fraction persists after inclusion of NNLO corrections both
with and without the loop-induced channel. In contrast to DPA, NWA result
overshoots the off-shell result by 1\%.  In both cases the differences to
the complete off-shell are well within their expectation of $\order{\Gamma_W/M_W}$,
even though this estimate is only exact for inclusive phase space integration. The scale
uncertainty decreases by a factor of 3 with NNLO corrections, however after
introduction of the loop-induced channel bounces back to 80\% and 50\% of the
NLO level for the higher and lower band correspondingly. The uncertainty
band of the LI channel is almost identical across setups, but its relative contribution
differs, affecting the overall theoretical uncertainty of the NNLO+LI result. Higher order
corrections to the loop-induced channel, which are formally of $\order{\as^3}$
(part of N${}^3$LO), are expected to improve the uncertainty
but are left for future work.

\begin{table}[!t]
  {\small
\begin{center}
\renewcommand{\arraystretch}{1.3}
\begin{tabular}{|C{2.2cm}C{2.26cm}C{2.26cm}C{1.35cm}C{2.26cm}C{2.26cm}|}%
\hline %
      \cellcolor{blue!9}
    & \cellcolor{blue!9}{NLO}
    & \cellcolor{blue!9}{NNLO}
    & \cellcolor{blue!9}{$K_{NNLO}$}
    & \cellcolor{blue!9}{LI}
    & \cellcolor{blue!9}{NNLO+LI}\\
\hline %
off-shell
    & $220.06(5)^{+1.8\%}_{-2.3\%}$
    & $225.4(4)^{+0.6\%}_{-0.6\%}$
    & 1.024
    & $13.8(2)^{+25.5\%}_{-18.7\%}$
    & $239.1(4)^{+1.5\%}_{-1.2\%}$ \\ 
unpol. (nwa)
    & $221.85(8)^{+1.8\%}_{-2.3\%}$
    & $227.3(6)^{+0.6\%}_{-0.6\%}$
    & 1.025
    & $13.68(3)^{+25.5\%}_{-18.7\%}$
    & $241.0(6)^{+1.5\%}_{-1.1\%}$ \\
unpol. (dpa)
    & $214.55(7)^{+1.8\%}_{-2.3\%}$
    & $219.4(4)^{+0.6\%}_{-0.6\%}$
    & 1.023
    & $13.28(3)^{+25.5\%}_{-18.7\%}$
    & $232.7(4)^{+1.4\%}_{-1.1\%}$ \\
\hline %
$W^+_L$ (dpa)
    & $57.48(3)^{+1.9\%}_{-2.6\%}$
    & $59.3(2)^{+0.7\%}_{-0.7\%}$
    & 1.032
    & $2.478(6)^{+25.5\%}_{-18.3\%}$
    & $61.8(2)^{+1.0\%}_{-0.8\%}$ \\
$W^-_L$ (dpa)
    & $63.69(5)^{+1.9\%}_{-2.6\%}$
    & $65.4(3)^{+0.8\%}_{-0.8\%}$
    & 1.026
    & $2.488(6)^{+25.5\%}_{-18.3\%}$
    & $67.9(3)^{+0.9\%}_{-0.8\%}$ \\
$W^+_T$ (dpa)
    & $152.58(9)^{+1.7\%}_{-2.1\%}$
    & $155.7(6)^{+0.7\%}_{-0.6\%}$
    & 1.020
    & $11.19(2)^{+25.5\%}_{-18.8\%}$
    & $166.9(6)^{+1.6\%}_{-1.3\%}$ \\
$W^-_T$ (dpa)
    & $156.41(7)^{+1.7\%}_{-2.1\%}$
    & $159.7(6)^{+0.5\%}_{-0.6\%}$
    & 1.021
    & $11.19(2)^{+25.5\%}_{-18.8\%}$
    & $170.9(6)^{+1.7\%}_{-1.3\%}$ \\
\hline %
$W^+_LW^-_L$ (dpa)
    & $9.064(6)^{+3.0\%}_{-3.0\%}$
    & $9.88(3)^{+1.3\%}_{-1.3\%}$
    & 1.090
    & $0.695(2)^{+25.5\%}_{-18.8\%}$
    & $10.57(3)^{+2.9\%}_{-2.4\%}$ \\
$W^+_LW^-_T$ (dpa)
    & $48.34(3)^{+1.9\%}_{-2.5\%}$
    & $49.4(2)^{+0.9\%}_{-0.7\%}$
    & 1.021
    & $1.790(5)^{+25.5\%}_{-18.3\%}$
    & $51.2(2)^{+0.6\%}_{-0.8\%}$ \\
$W^+_TW^-_L$ (dpa)
    & $54.11(5)^{+1.9\%}_{-2.5\%}$
    & $55.5(4)^{+0.6\%}_{-0.7\%}$
    & 1.025
    & $1.774(5)^{+25.5\%}_{-18.3\%}$
    & $57.2(4)^{+0.7\%}_{-0.7\%}$ \\
$W^+_TW^-_T$ (dpa)
    & $106.26(4)^{+1.6\%}_{-1.9\%}$
    & $108.3(3)^{+0.5\%}_{-0.5\%}$
    & 1.019
    & $9.58(2)^{+25.5\%}_{-18.9\%}$
    & $117.9(3)^{+2.1\%}_{-1.6\%}$ \\
\hline %
\end{tabular}
\end{center}
\caption{Total cross-sections (in fb) for the unpolarised, singly-polarised and
  doubly-polarised $\PW^+\PW^-$ production at the LHC. Unpolarised calculation
  is performed in three ways: full off-shell, and using approximations: DPA,
  NWA. Polarised setups are calculated using DPA.
  Uncertainties are computed with 7-point scale variations by a factor of
  $2$ around the central scale. $K$-factors are presented as ratios of NNLO QCD
  (without LI) over NLO integrated cross-sections.
  NNLO+LI represents the full $\mathcal{O}(\alpha_s^2)$ result.
} \label{table:totalxsec} }
\end{table}

Next we consider singly-polarised setups. Missing interferences between
longitudinal and transverse polarisations of $\PW^+(\PW^-)$ and restrictions on
the leptonic phase space due to cuts, give rise to differences between the sum
of the singly-polarised setups and the fully unpolarised DPA setup.
The interference effects are of order 2\% and are negative (positive)
for singly-polarised $\PW^+$($\PW^-$).
The NNLO
K-factors are of the same magnitude across setups, with a slightly larger value
for the longitudinal polarisations. The scale uncertainty features the same
behaviour as for the unpolarised setups, except in the longitudinal setups it
is not amplified by the loop-induced channel as it has a smaller relative
contribution in comparison with transverse setups.

\begin{figure}[t]
  \centering
  \subfigure[at NLO]{\includegraphics[width=0.49\textwidth]
      {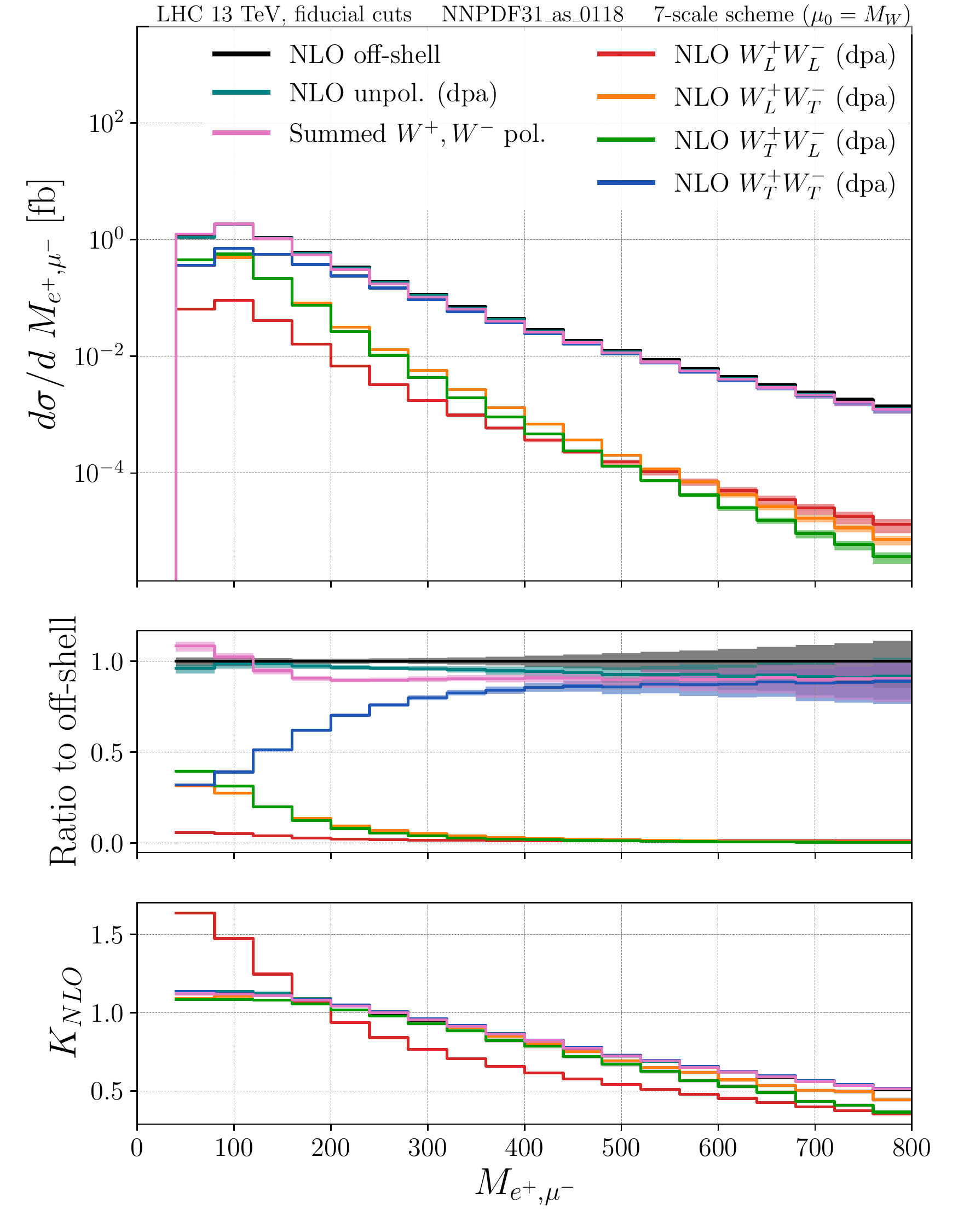}}
  \subfigure[at NNLO (without LI)]{\includegraphics[width=0.49\textwidth]
      {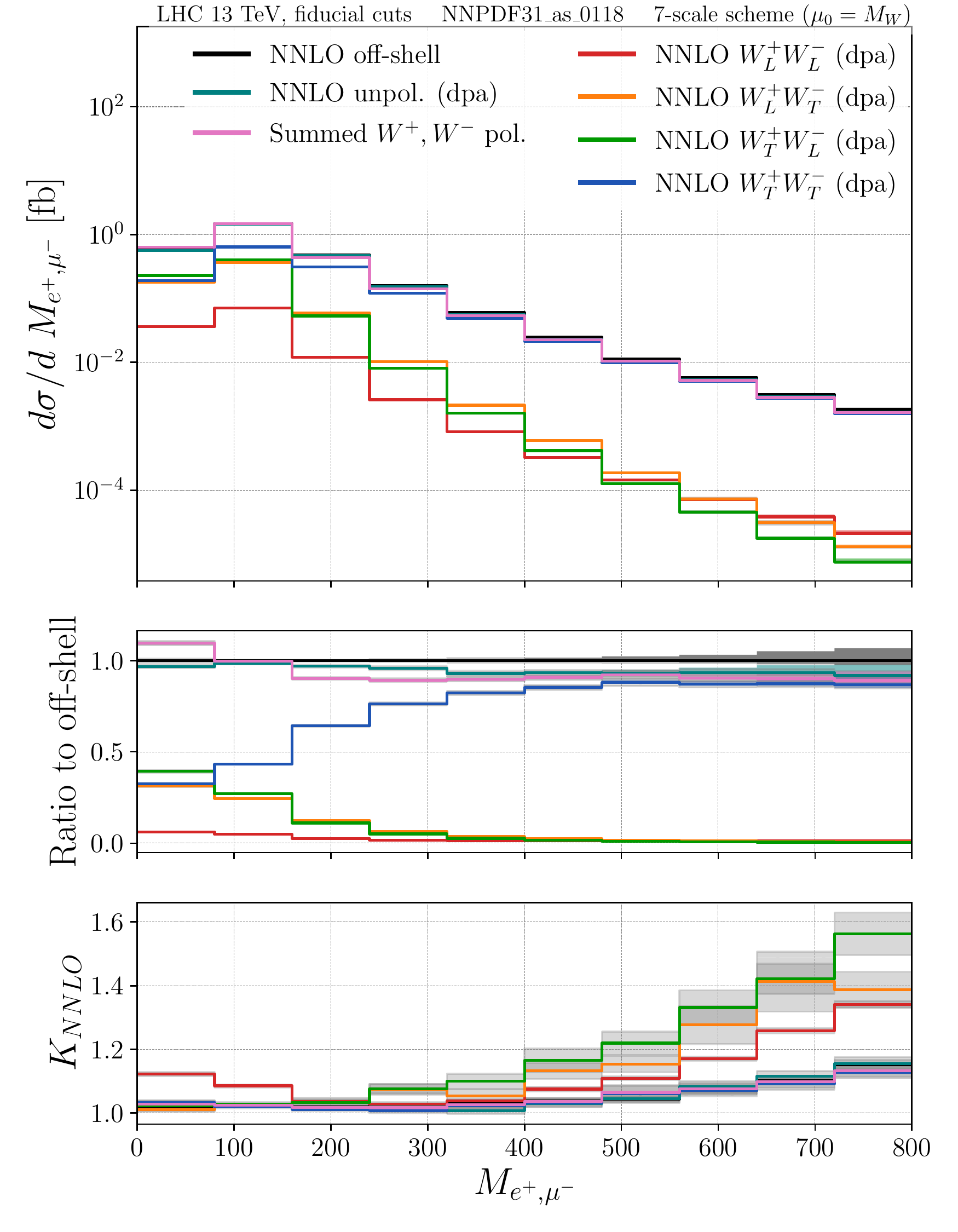}}
  \caption{
      Distribution of charged lepton pair invariant mass at different
      orders of QCD. Doubly-polarised setups are
      shown. From top-down: absolute value differential distribution,
      ratio to off-shell result, K-factor of the corresponding order.
      Monte-Carlo errors are shown in the middle and lower panes as grey bands.
      Scale variation bands are shown as coloured bands in the upper pane.
        \label{fig:m_ll} \label{fig:plot_description}}
\end{figure}

NNLO corrections (without LI) to the doubly-polarised setups can be estimated
to be roughly 25\% of NLO corrections, except for TL setup where NNLO
corrections are a bit larger. It was shown that at NLO the doubly-longitudinal
polarisation of the diboson system is, among other polarisation setups,
particularly affected by QCD corrections \cite{Denner:2020bcz}. This is also true
at NNLO as represented by the corresponding K-factor. As will be shown
further, the profile of its corrections is also distinctly different on the
differential level. The scale variation band goes down at NNLO, however the
loop-induced channel brings it back to NLO level at both LL and TT setups,
whereas for LT and TL setups it remains on the same level.

Of interest are the polarisation fractions, i.e. the fractions of the
cross section for various polarised boson configurations. Although NNLO
corrections differ among the different polarisations, there is no significant
difference in the polarisation fractions with respect to NLO, indicating
that the fractions are rather independent of higher order QCD corrections.
In particular, the fraction of doubly-longitudinal polarised $\PW$, which gets
the largest corrections, is still small.

\subsection{NNLO QCD corrections to differential cross sections}
\label{subsec:pure_nnlo_effects}

\begin{figure}[!ht]
  \centering
  \subfigure[at NLO]{\includegraphics[width=0.49\textwidth]
      {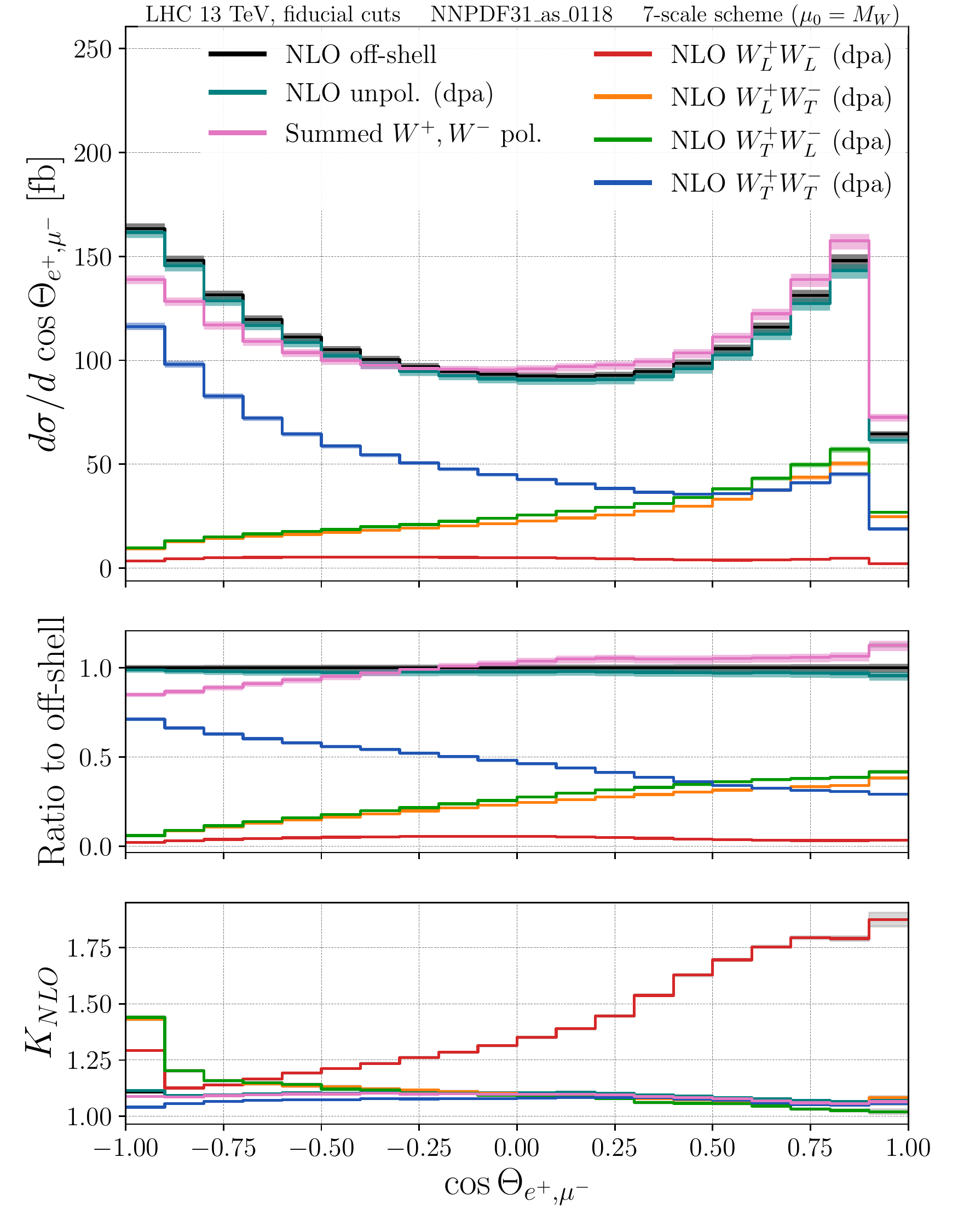}}
  \subfigure[at NNLO (without LI)]{\includegraphics[width=0.49\textwidth]
      {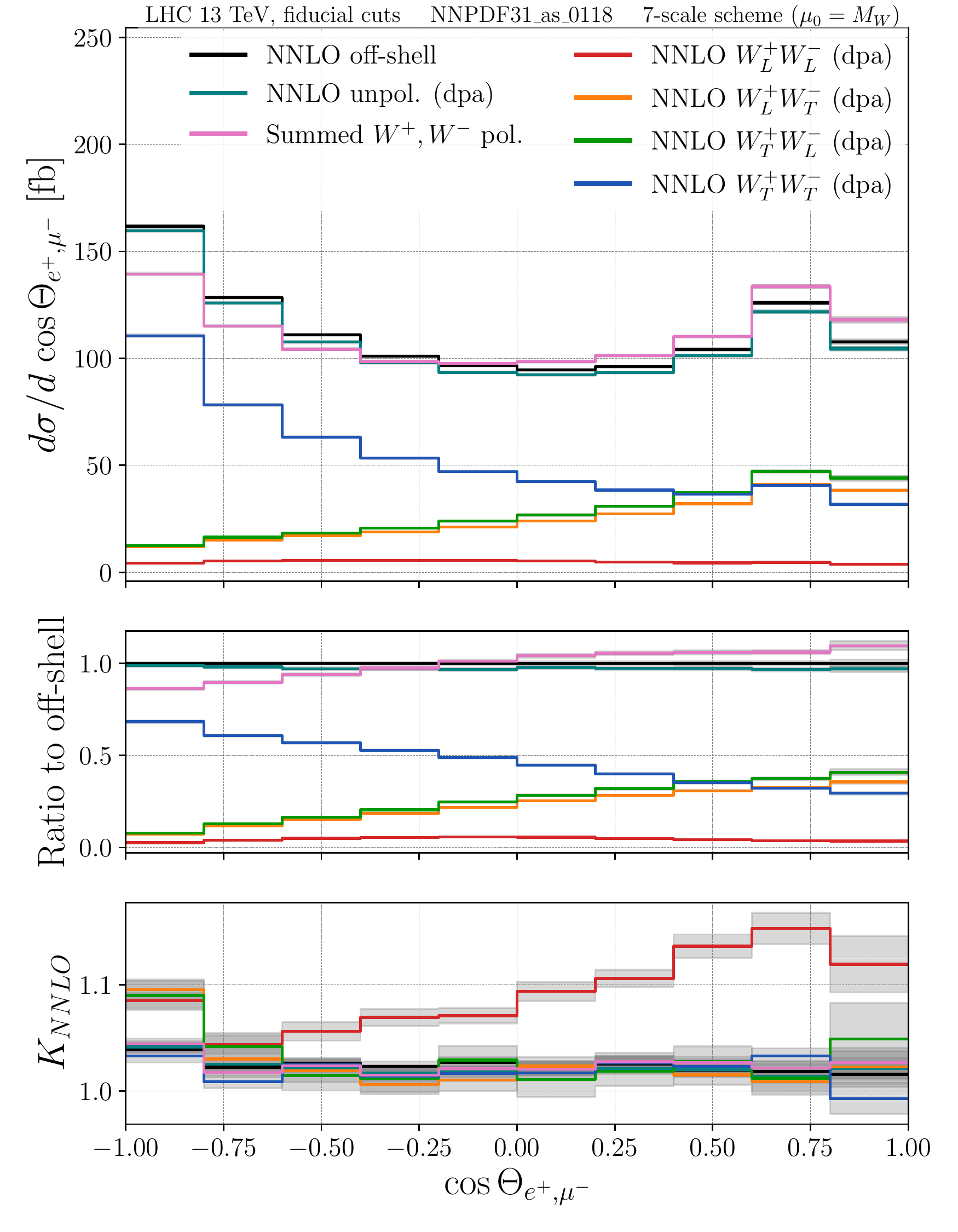}}
  \caption{
      Distribution of cosine of angle between charged leptons.
      Doubly-polarised setups are shown. Individual plot substructure
      is the same as in \reffi{fig:plot_description}.
        \label{fig:costheta_lep_SS}}
\end{figure}

In this section we will explore NNLO QCD effects on the differential distributions
as they appear without the loop-induced channel. Observables which allow
discrimination between different boson polarisations are of particular interest,
theoretically and experimentally. The key quantity here is again the
(differential) polarisation fractions.  A general feature of differential
polarisation fractions is that at high energies the longitudinal component
vanishes as the weak bosons get effectively massless. Naturally, regions of large
invariant mass or transverse momentum are populated by transversely polarised
$\PW$-bosons. Close to the $\PW$-pair production threshold, where the diboson system is
produced with small momentum, the contribution of the longitudinal component is
largest.

This characteristic can be seen in the invariant mass distribution of the
charged lepton pair, shown in \reffi{fig:m_ll}. We show the NLO (left) and NNLO
(right) predictions for the absolute cross section, the differential
polarisation fraction, and the NLO and NNLO K-factors, respectively. At NNLO, the tail
features strong positive corrections for the setups with longitudinally
polarised bosons, reaching 50\%. The scale variation bands get notably
reduced across polarisation setups, particularly in the tail of the
distribution, however there the off-shell calculation still shows a substantial
scale uncertainty.
The shape of the NNLO K-factor in the tail is mainly driven
by the fixed scale choice which is not optimal for this phase space region.
Finally, the doubly-longitudinal setup shows a larger
correction in the low invariant mass region, which was also the case at NLO.

\begin{figure}[t]
  \centering
  \subfigure[at NLO]{\includegraphics[width=0.49\textwidth]
      {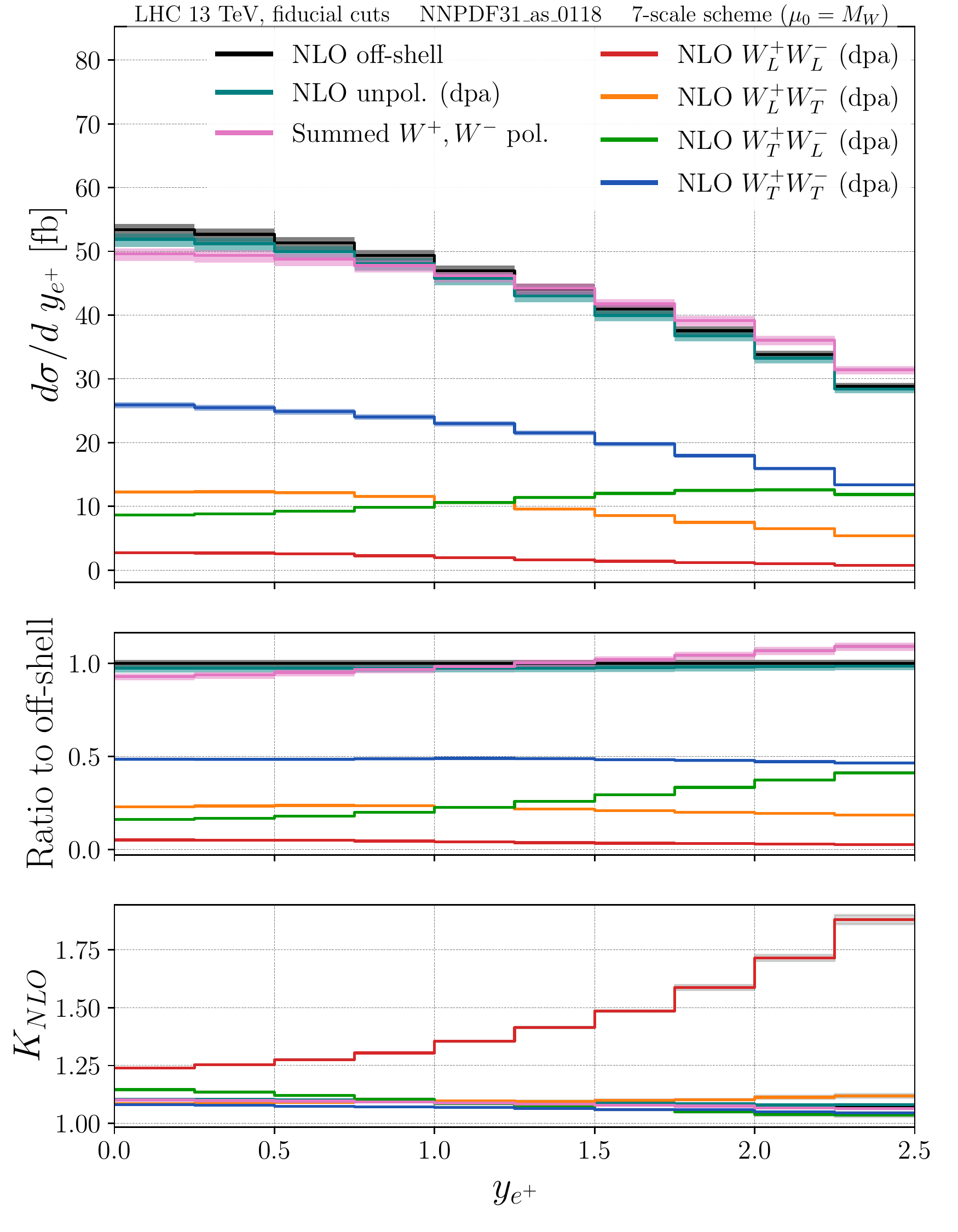}}
  \subfigure[at NNLO (without LI)]{\includegraphics[width=0.49\textwidth]
      {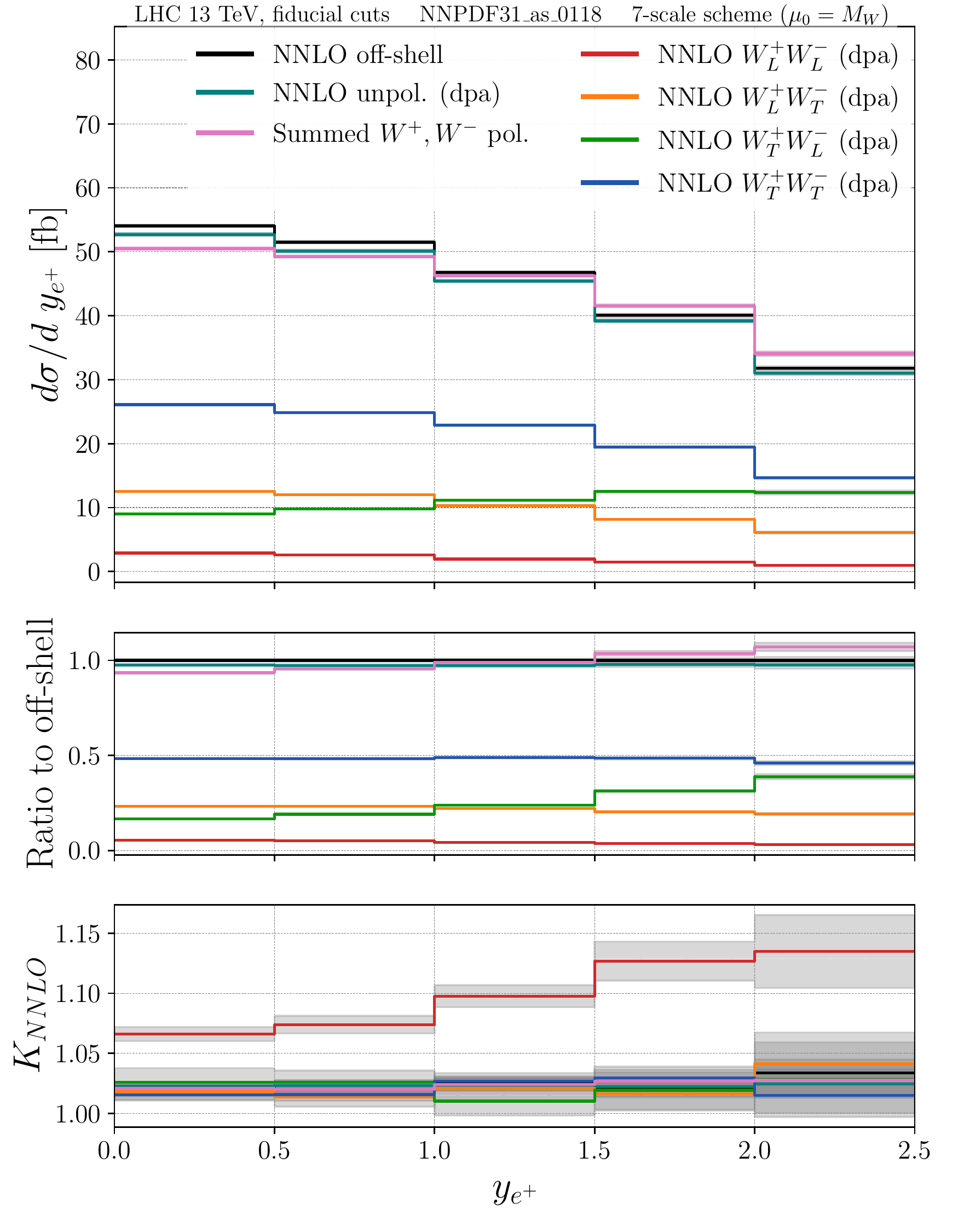}}
  \caption{
      Symmetrised distribution of $\Pe^+$ rapidity at different orders of QCD.
      Doubly-polarised setups are shown. Individual plot substructure is the
      same as in \reffi{fig:plot_description}.
        \label{fig:rapidity-ep}}
\end{figure}

It is worth pointing out that this is the region where the most of the production
cross section is coming from. This implies that observables which are sensitive
to the threshold region, or bulk region, are especially well suited to study
boson polarisations. In particular, this includes angular observables of the
final state charged leptons which both have a strong sensitivity to
polarisations and are shaped by the bulk region.

For example, consider the angular separation between two charged leptons in
\reffi{fig:costheta_lep_SS}. Back-to-back configurations, i.e.  where
$\cos\Theta_{\Pe^+,\mu^-} \approx -1$, are dominated by the doubly-transverse setup,
while the regions where the two leptons are aligned, have large contributions
from setups containing a longitudinal boson. NNLO corrections reduce the scale
dependence to the sub-percent level and show a rather small and flat K-factor.
A notable exception is the LL setup as it receives strong corrections up to
$10-15\%$ in magnitude and shape.  However, due to its overall small
contribution it does not affect the polarisation fractions.

Similar effects can be observed in the rapidity distributions.
\reffi{fig:rapidity-ep} features the symmetrised version of $\Pe^+$ rapidity
distribution. Here, LL polarisation receives a significant correction for higher
rapidities.

\begin{figure}[!ht]
  \centering
  \subfigure[Positron transverse momentum]
    {\includegraphics[width=0.48\textwidth]
      {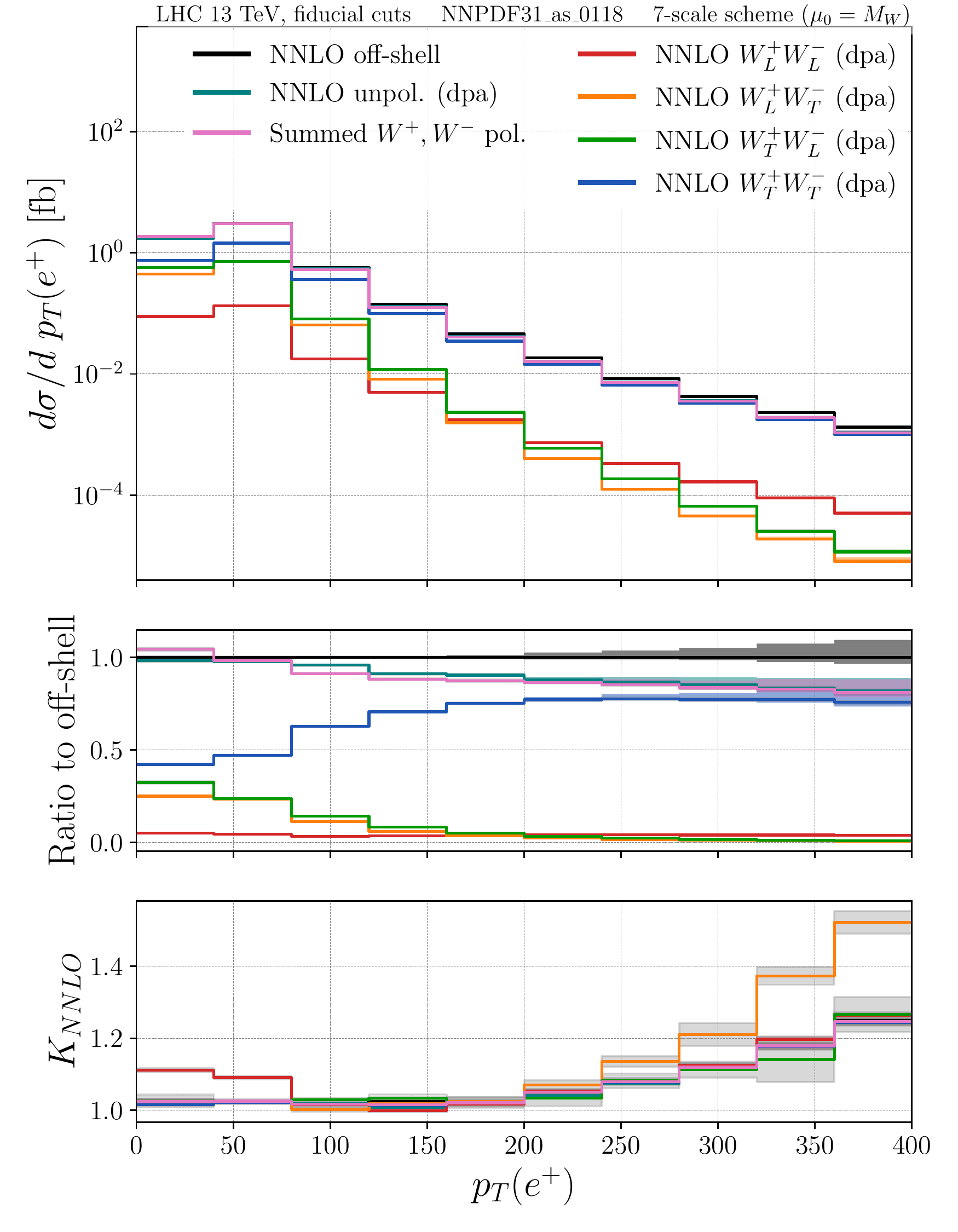}\label{fig:pTlep-positron}}
  \subfigure[Ratio of the muon and positron transverse momentum distributions]{
    \includegraphics[width=0.48\textwidth]
      {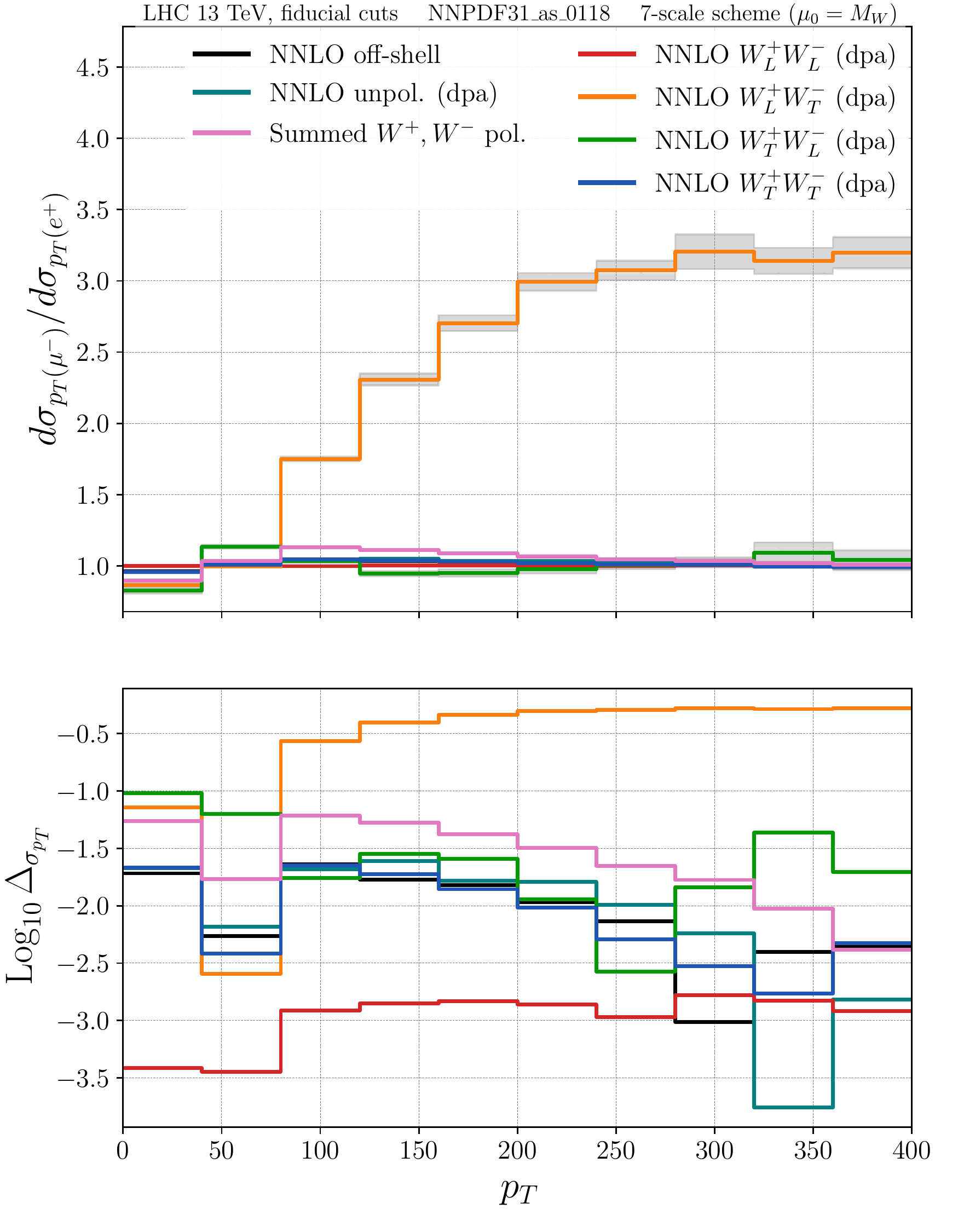}\label{fig:pTlep-diff}}

  \subfigure[$\cos\angle(\Pe^+,\PW^+)$]
      {\includegraphics[width=0.48\textwidth]
      {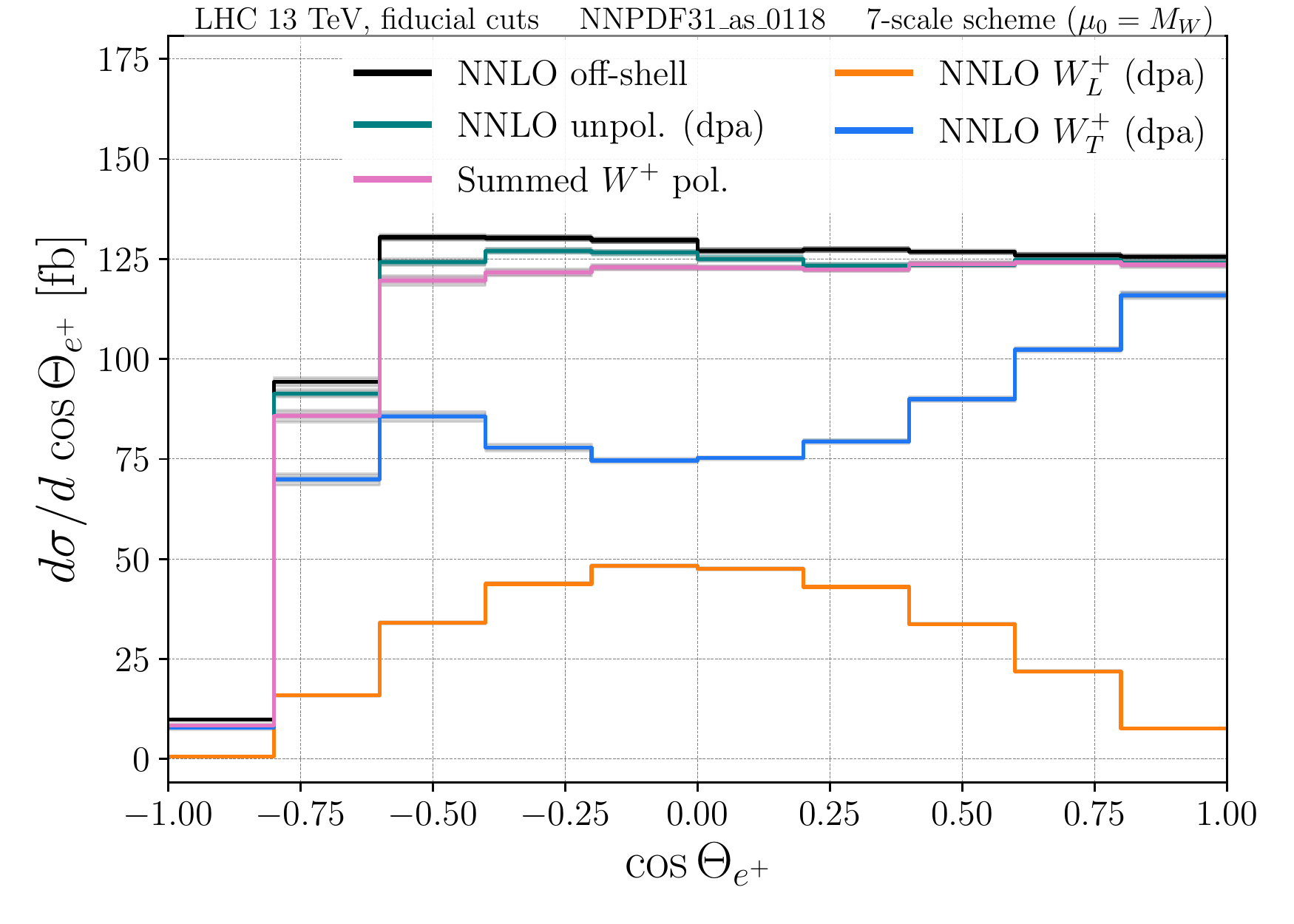} \label{fig:cosTheta-positron}}
  \subfigure[$\cos\angle(\mu^-,\PW^-)$]
      {\includegraphics[width=0.48\textwidth]
      {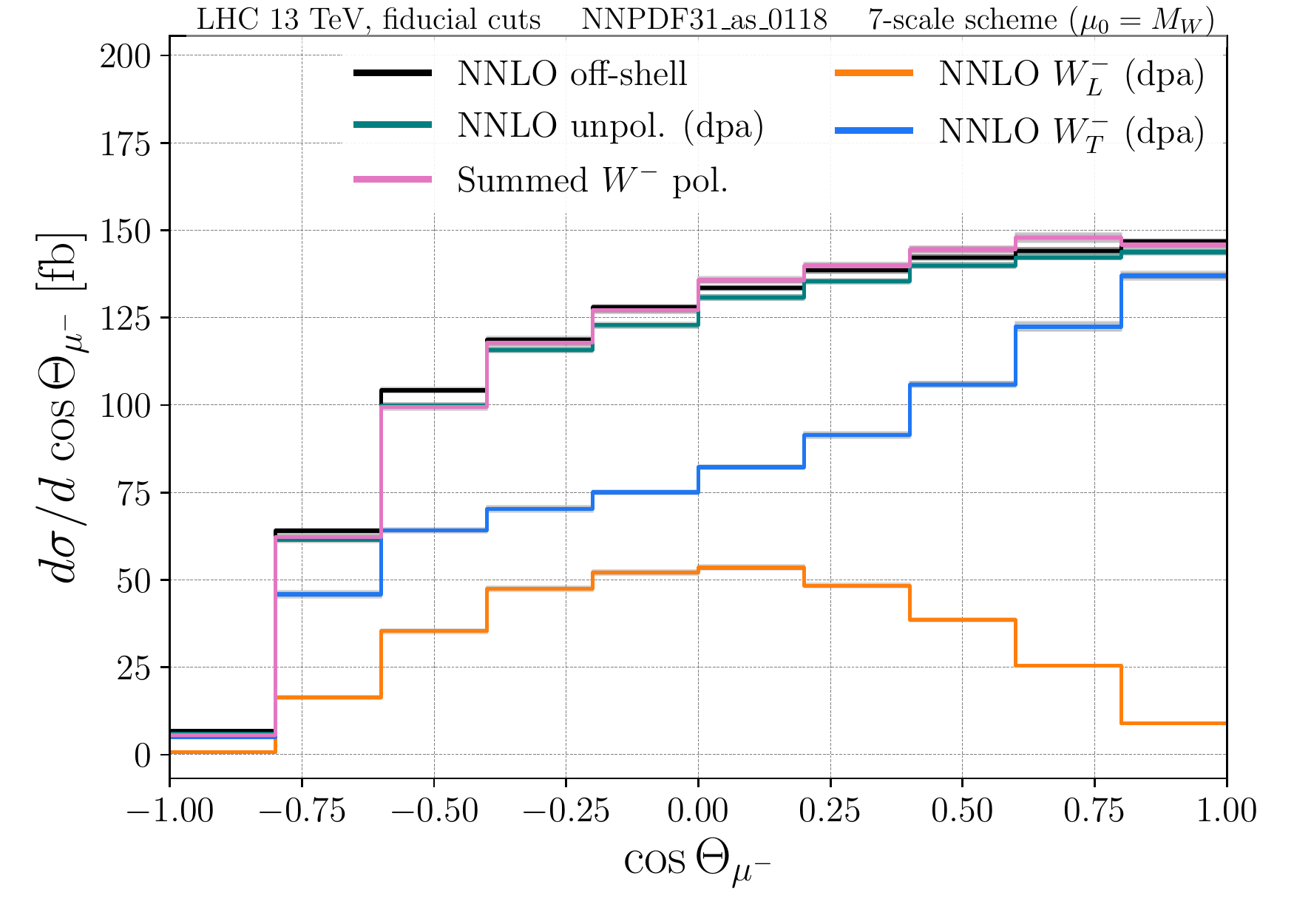} \label{fig:cosTheta-muon}}
    \caption{
        (Upper row) Distribution of lepton transverse momentum at NNLO (without LI)
        for $\Pe^+$ (left) and its comparison with distribution for $\mu^-$ (right).
        Individual plot substructure in (a) is the same as in
        \reffi{fig:plot_description}. The plot (b) structure is the
        following: the upper plot features the ratio of muon over positron
        transverse momenta distributions; the lower plot in (b) features
        $\Log_{10}\left|\frac{\sigma_{p_T}(\Pe^+) -
            \sigma_{p_T}(\mu^-)} {\sigma_{p_T}(\Pe^+) +
            \sigma_{p_T}(\mu^-)}\right|$, where $\sigma_{p_T}$ is the
        differential transverse momentum distribution. Doubly-polarised setups are
        shown.
        (Lower row) Distributions of charged lepton scattering angle cosine
        calculated in the parent $\PW$-boson CM frame at NNLO (without LI) for
        $\Pe^+$ (left) and $\mu^-$ (right). Parent boson singly-polarised setups
         are shown. Individual plots show the absolute value distributions.
  \label{fig:pTlep-both}}
\end{figure}

To point out another feature of the longitudinally polarised signals, it is
instructive to investigate transverse momentum distributions for leptons and
$\PW$-bosons. First, consider the transverse momentum distributions of the
charged leptons $p_T(\Pe^+)$ and $p_T(\mu^-)$. In \reffi{fig:pTlep-both} we
show $p_T(\Pe^+)$ (left) and the ratio of the differential cross sections
(right)
\begin{equation}
 \frac{\dd \sigma_{p_T(\mu^-)}}{\dd \sigma_{p_T(\Pe^+)}} \equiv
 \frac{\dd \sigma/\dd p_T(\mu^-)}{\dd \sigma / \dd p_T(\Pe^+)}\;.
\end{equation}
There is a striking difference in the ratio for the LT polarised setup, which can
be explained through asymmetries in the decay of $\PW^+$ and $\PW^-$. The
charged leptons have a larger probability to get emitted forward (in the
flight direction of the parent $\PW$ boson) for transversely polarised bosons
than for longitudinal ones. This can be seen in \reffi{fig:cosTheta-positron}
and \reffi{fig:cosTheta-muon} demonstrating the distribution in the opening
angle between the charged leptons and their parent $\PW$-boson (for precise
definition of the angles, see \refapp{sec:appendix_azimuth}).
Thus we expect the transverse boson to produce a harder $p_T$ spectrum
for its decay products than the longitudinal one, i.e. the ratio $\dd
\sigma_{p_T(\mu^-)}/\dd \sigma_{p_T(\Pe^+)}$ is expected to be smaller than $1$
for TL setup and larger than $1$ for LT setup.
This effect is magnified by the asymmetry between angle of emission distribution
in $\PW^+$ and $\PW^-$ transverse setups and is caused by the universally
left-handed nature of $\PW$-bosons produced at the LHC \cite{Bern:2011ie}.
Unfortunately, as in the case of the invariant mass distribution, the
longitudinal contributions vanish at
high transverse momentum, which makes it more difficult to exploit this
observation in experimental measurements.

\begin{figure}[!htb]
  \centering
  \includegraphics[width=0.99\textwidth]{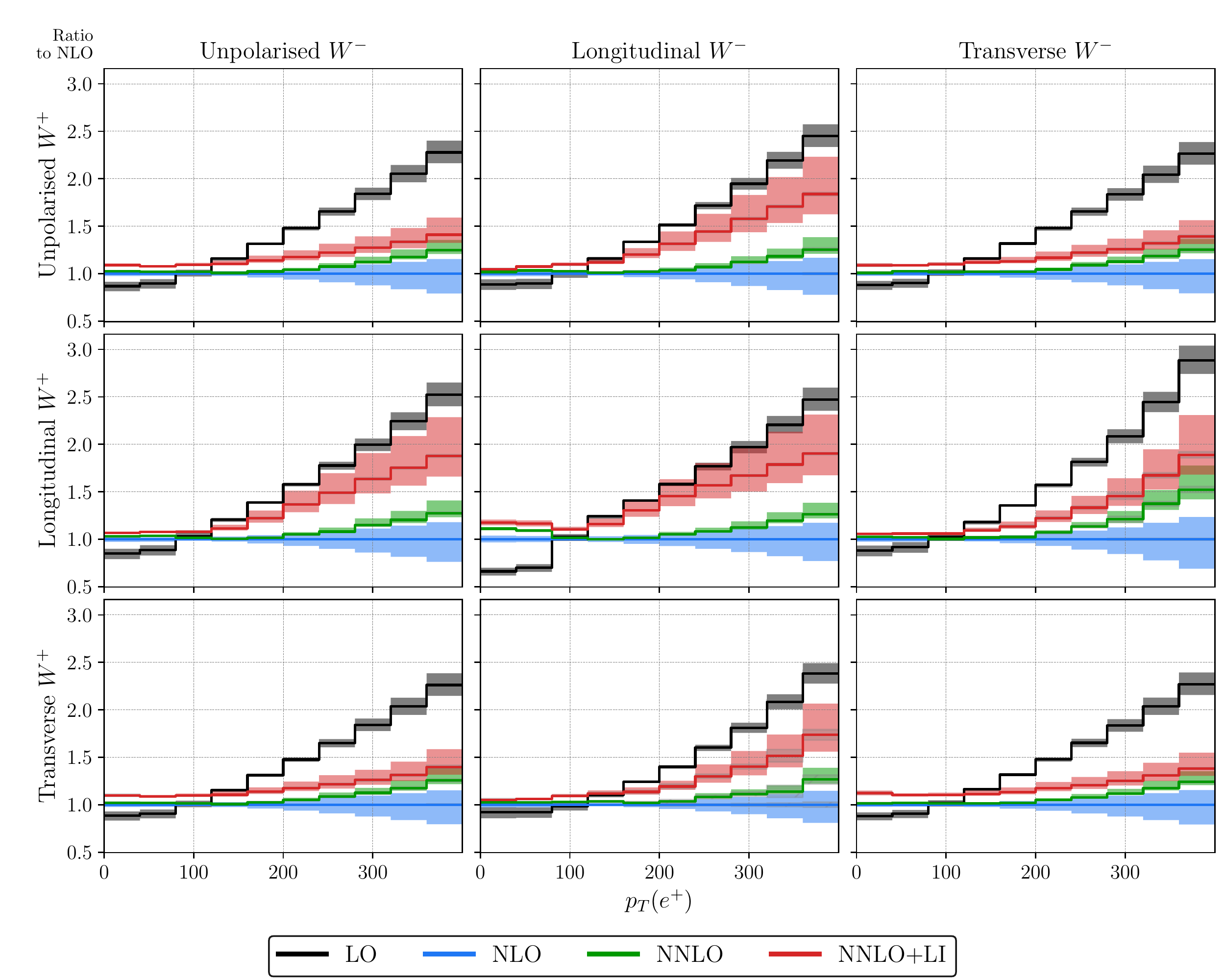}
  \caption{Ratio to NLO of $\Pe^+$ transverse momentum distribution at various
    orders of perturbative QCD. Each of 9 panes represents a selected
    polarisation setup calculated within DPA framework.  $\PW^+$ and $\PW^-$
    polarisation setups $\{U,L,T\}$ are cycled across vertical and horizontal
    plots respectively. Coloured regions represent scale variation bands, and grey
    bands -- Monte-Carlo uncertainties.
  \label{fig:pT_electron_setups}}
\end{figure}

\subsection{Effects of the loop-induced contribution}
    \label{subsec:effects_of_loop_induced_contribution}

The loop-induced $gg$-channel only appears at $\as^2$ order and has quite a
substantial effect on the cross-section. Its effects on various differential
distributions have been already investigated in \cite{Denner:2020bcz}. In this
section we will briefly comment on it again in the context of the NNLO
calculation.

Usually, the loop-induced channel provides a glimpse into the NNLO effects in
general. However, due to the simple kinematics structure of the diboson
production, especially its double-resonant contribution, the Born kinematics
configuration is not enough to get the right distribution shape for many
observables. Using \refta{table:totalxsec} it can be pointed out that
loop-induced channel increases scale variation bands up to almost NLO QCD level
in all setups except the ones that contain exactly one longitudinally polarised
weak boson.

\begin{figure}[!htb]
  \centering
  \includegraphics[width=0.99\textwidth]{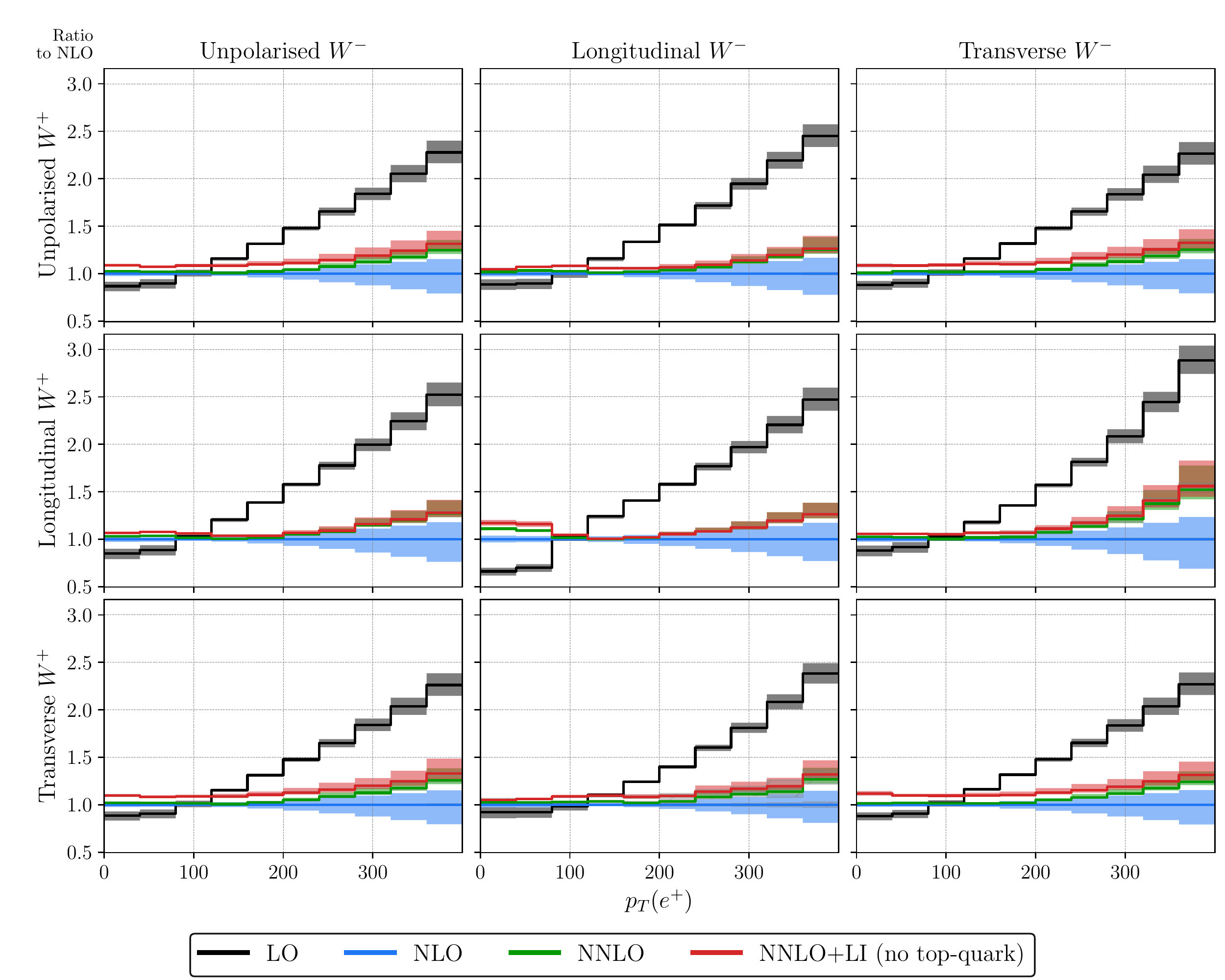}
  \caption{
    Ratio to NLO of $\Pe^+$ transverse momentum distribution at various
    orders of perturbative QCD with top-quark loop contribution removed.
    Same plot structure as in \reffi{fig:pT_electron_setups}.
  \label{fig:pT_electron_setups_nf5}}
\end{figure}

In \reffi{fig:pT_electron_setups} we present relative corrections to the
differential distribution of positron transverse momentum, with respect to the
NLO calculation. Here we observe that NNLO calculation has brought the scale
variation down, and that it is within a reasonable distance from NLO given the scale
uncertainty. However, the loop-induced contribution drastically changes the
picture, and its effect depends on the polarisation setup.

Setups that include a longitudinal $\PW$-boson receive mild corrections from the
loop-induced channel at low $p_T$ but are hugely affected in the tail ($p_T > 120
\GeV$). Also, the scale variation band in the tail becomes larger than for any
other approximation.
As is illustrated by \reffi{fig:pT_electron_setups_nf5}, this is caused mainly by the top-quark
loop contribution, which becomes
relevant at $p_T(l^\pm) \sim 100 \GeV$. It does not affect the results or the scale uncertainty
at the total cross section level ant it is not visible in any of the distribution
representing the bulk of the cross section.

\begin{figure}[!ht]
  \centering
  \includegraphics[width=0.99\textwidth]{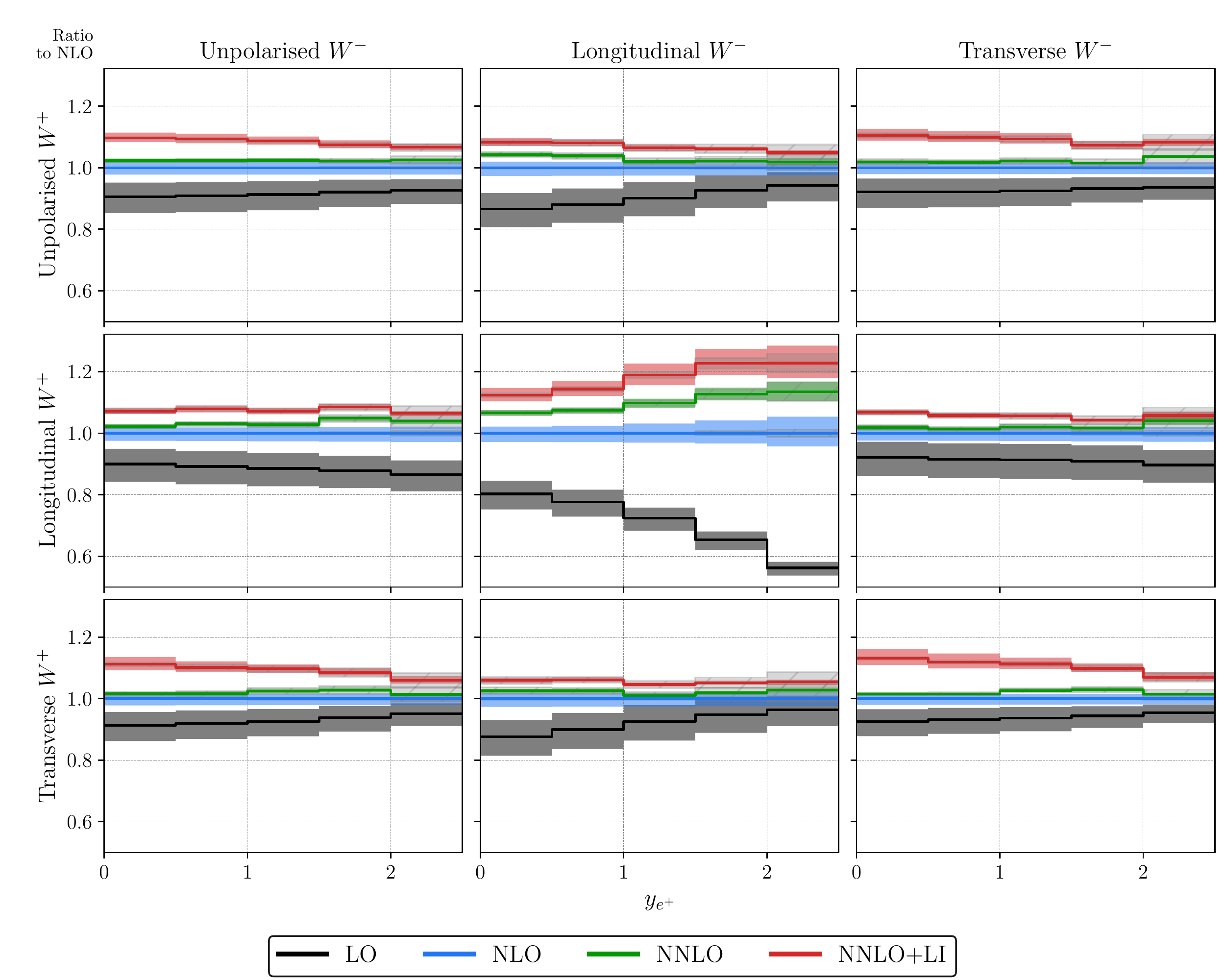}
  \caption{
    Ratio to NLO of symmetrised $\Pe^+$ rapidity distribution at various orders of
    perturbative QCD.  Same plot structure as in \reffi{fig:pT_electron_setups}.
      \label{fig:rapidity_ep}}
\end{figure}

Setups containing a transverse $\PW$-boson represent a different correction
in the loop-induced channel at NNLO. Here we observe an overall positive shift
of order 10\% which diminishes by the end of the distribution. This behaviour is
expectedly replicated by the unpolarised setup as the transverse
contribution dominates the cross section.

Similar effects can be observed in the charged leptons invariant mass
distribution.

Next, we show the rapidity of $\Pe^+$ distribution in
\reffi{fig:rapidity_ep}. The setups containing transverse polarisations as well
as the unpolarised setup feature about a 10\% positive correction which is
compatible with what we observed at the bulk of the positron $p_T$ distribution.
The case is different however for setups containing a longitudinal $\PW$. In particular,
the LL polarisation receives a strong correction, which is expected due to its
overall large K-factor in \refta{table:totalxsec}. The $gg$ loop-induced channel
follows the shape of NNLO corrections except in the TT setup, where the loop
induced corrections appear to be larger than NLO.

\begin{figure}[!htb]
  \centering
  \includegraphics[width=0.99\textwidth]{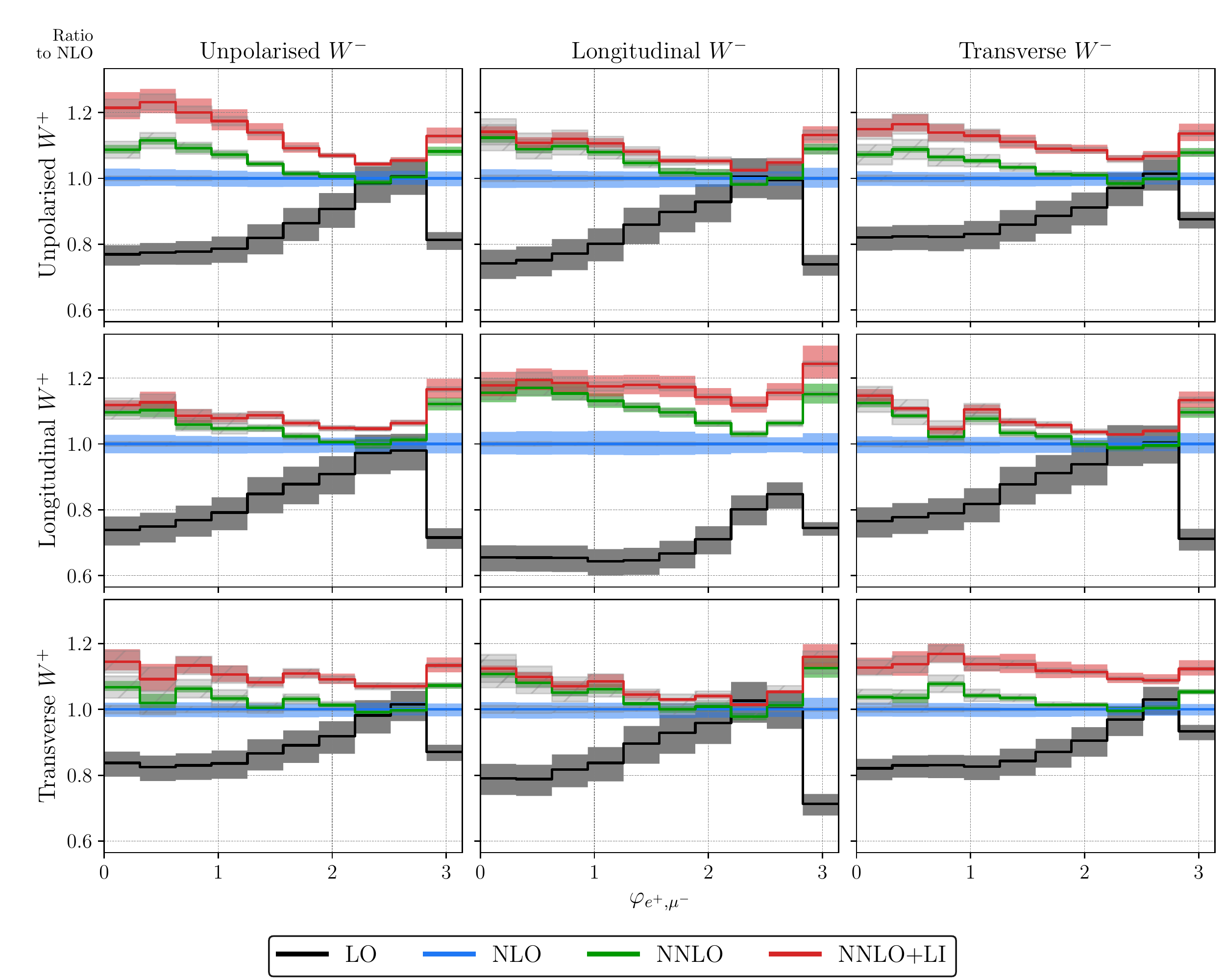}
  \caption{
    Ratio to NLO of azimuthal separation between charged leptons at various orders of
    perturbative QCD. Same plot structure as in \reffi{fig:pT_electron_setups}.
  \label{fig:azimuth-lep}}
\end{figure}

Finally, in angular distributions related to the lepton emission angles we see an overall
shift which does not affect the distribution shapes. However, a notable difference can
be observed in azimuthal separation between the charged leptons which is a
distribution highly susceptible to interference effects even at the inclusive level.
In \reffi{fig:azimuth-lep} we note that LI channel has a large overall shift in TT
and unpolarised setups and features a rather interesting behaviour in LL setup.
LI channel barely has any effect on LL setup at $\phi_{\Pe^+,\mu^-} < 1$ but
then introduces sizeable corrections up to 15\%.  We will encounter a similar
shape in the discussion about DPA and NWA at unpolarised level in
\refse{subsec:comparison_between_dpa_and_nwa}.

Overall, we see a need for higher corrections of order $\as^3$
in the $gg$ loop-induced channel, to bring the scale variation down. This is left
for future work.

\subsection{Comparison between DPA and NWA}\label{subsec:comparison_between_dpa_and_nwa}

\begin{figure}[!ht]
  \centering
  \subfigure[Symmetrised rapidity of positron (NLO)]
    {\includegraphics[width=0.48\textwidth]
      {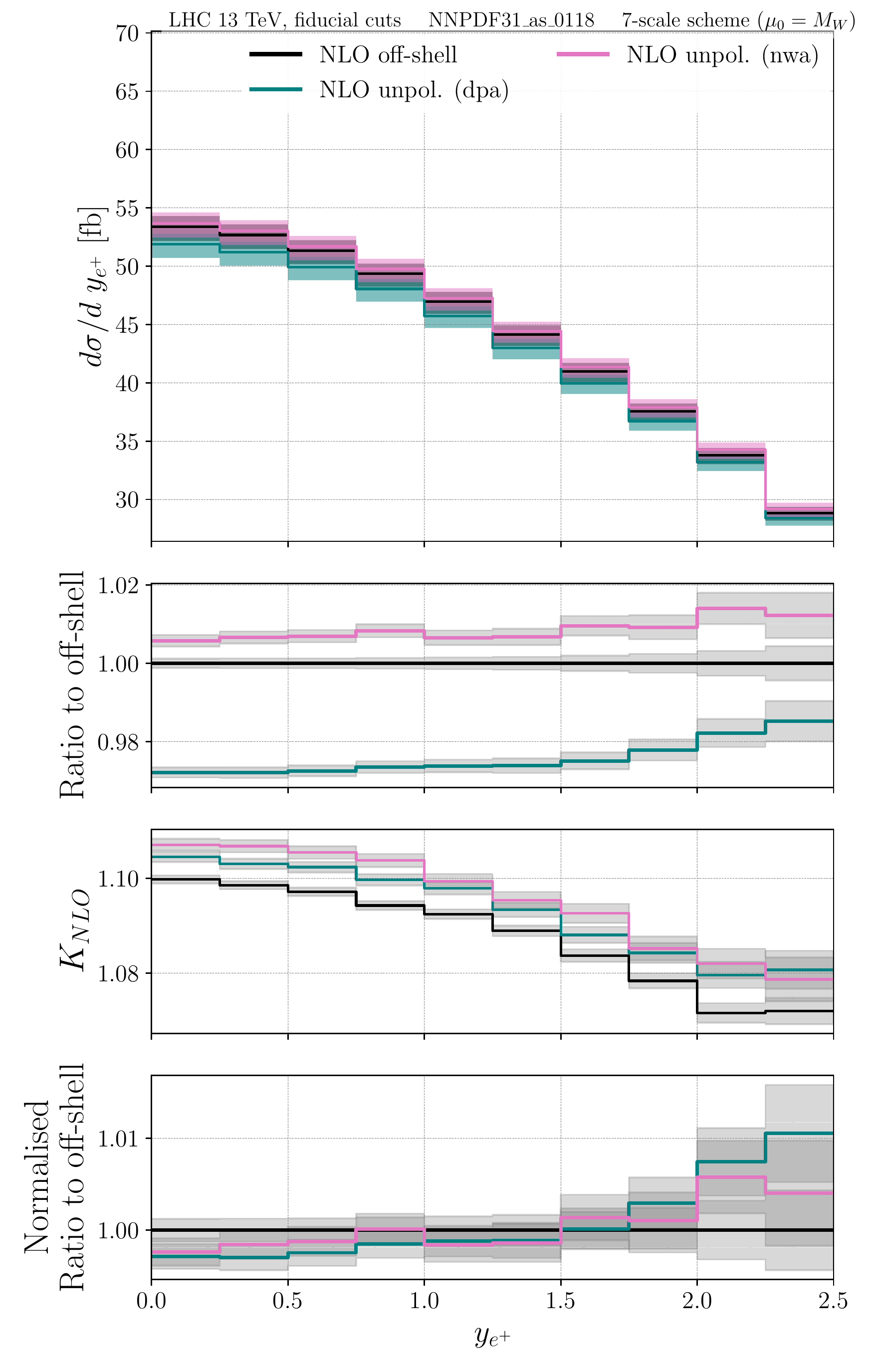}
        \label{fig:setups_rapidity}}
  \subfigure[Symmetrised azimuthal angle of positron emission in its parent boson CM (NLO)]
    {\includegraphics[width=0.48\textwidth]
      {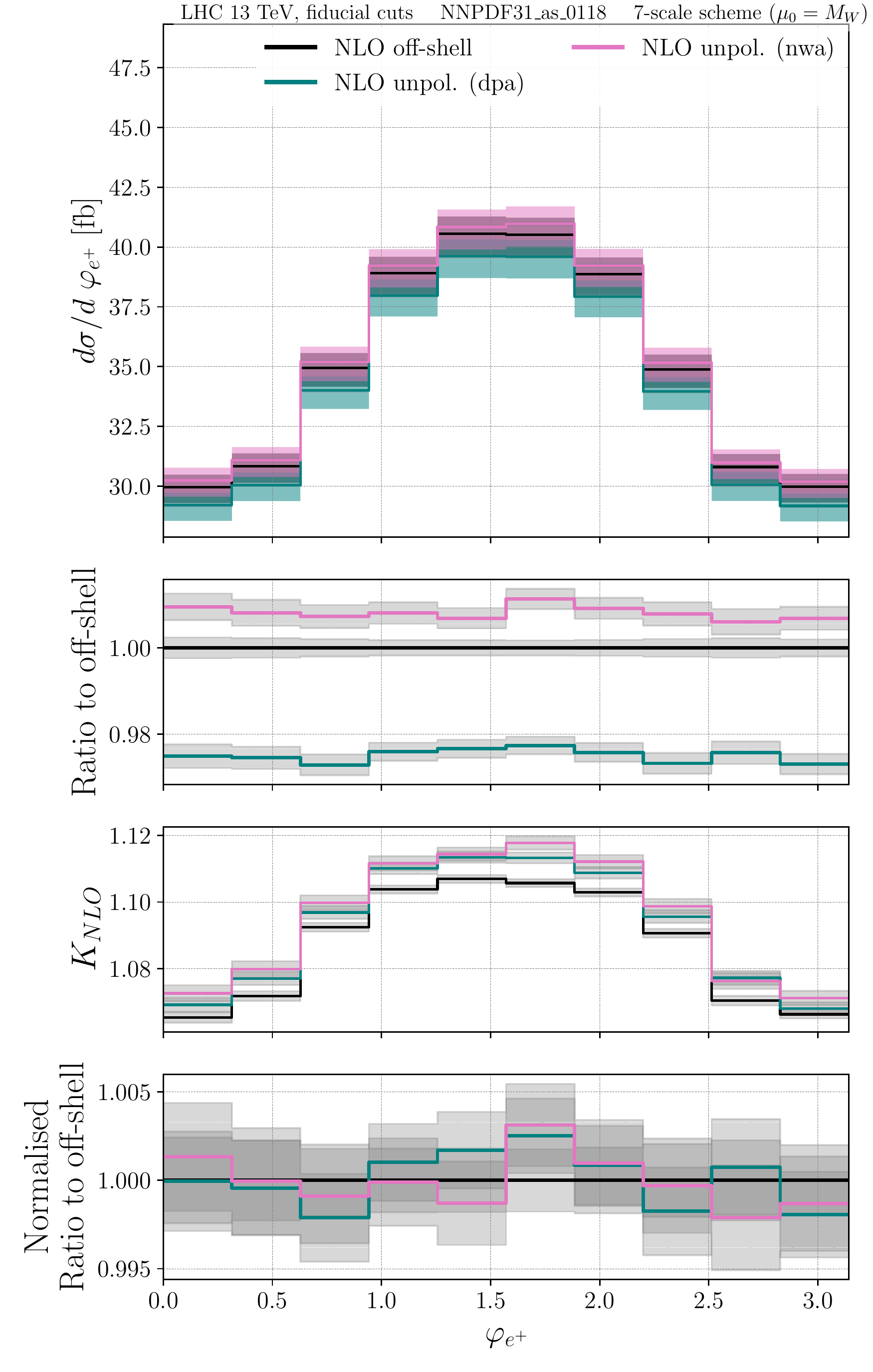}
      \label{fig:setups_azimuthEmission}}
  \caption{
      Comparison between off-shell calculation, DPA, and NWA for selected
      distributions. Top three panes have the same structure as in
      \reffi{fig:plot_description}; the bottom plot shows the ratio
      distributions to off-shell calculation normalised according to integrated
      cross-section value.
        \label{fig:setups_similarities}}
\end{figure}

\begin{figure}[!htb]
  \centering
  \subfigure[Cosine of positron emission polar angle]
    {\includegraphics[width=0.48\textwidth]
      {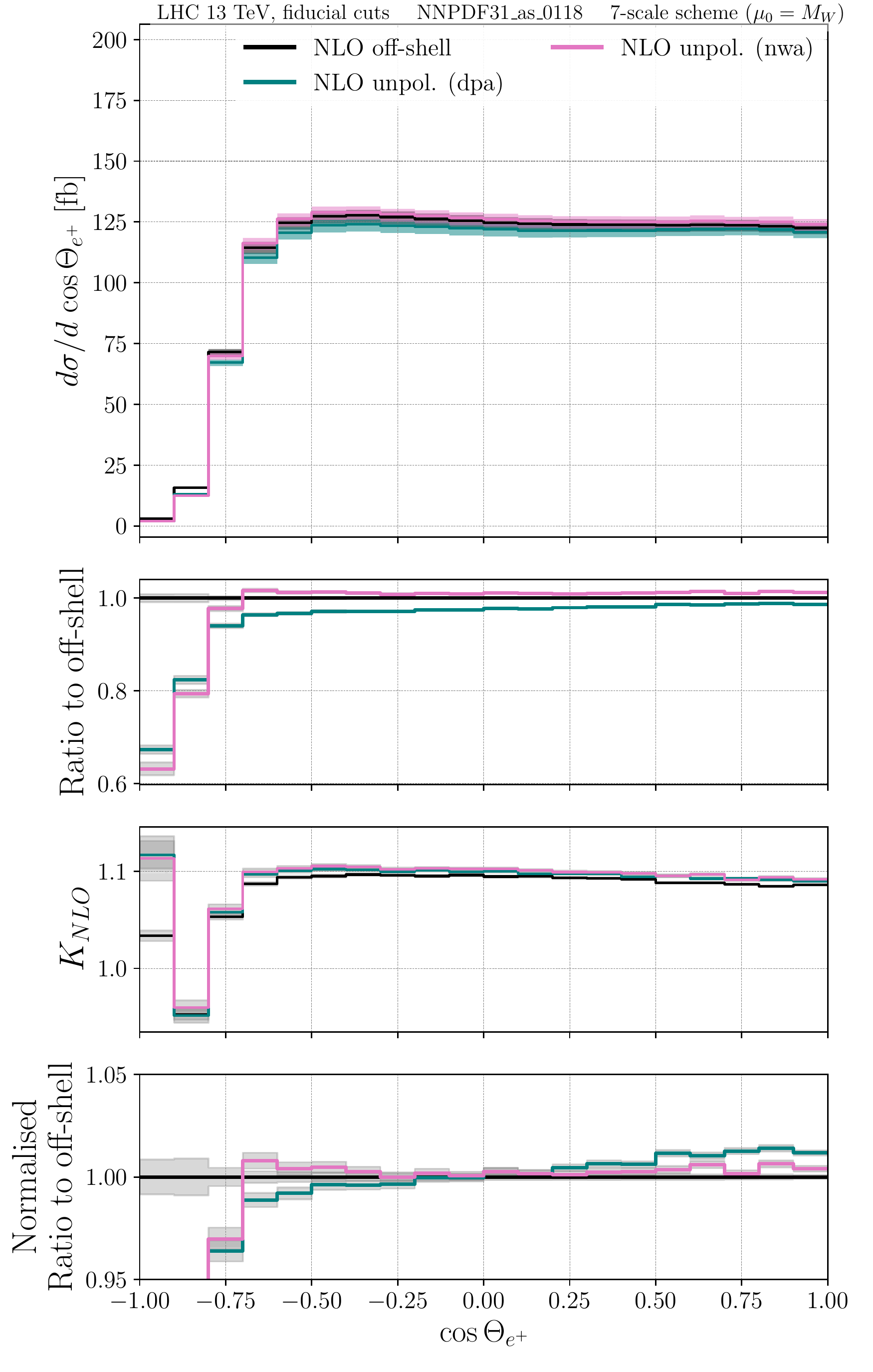}
      \label{fig:setups_cosThetaEmission}}
  \subfigure[Cosine of angle between charged leptons (NLO)]
    {\includegraphics[width=0.48\textwidth]
      {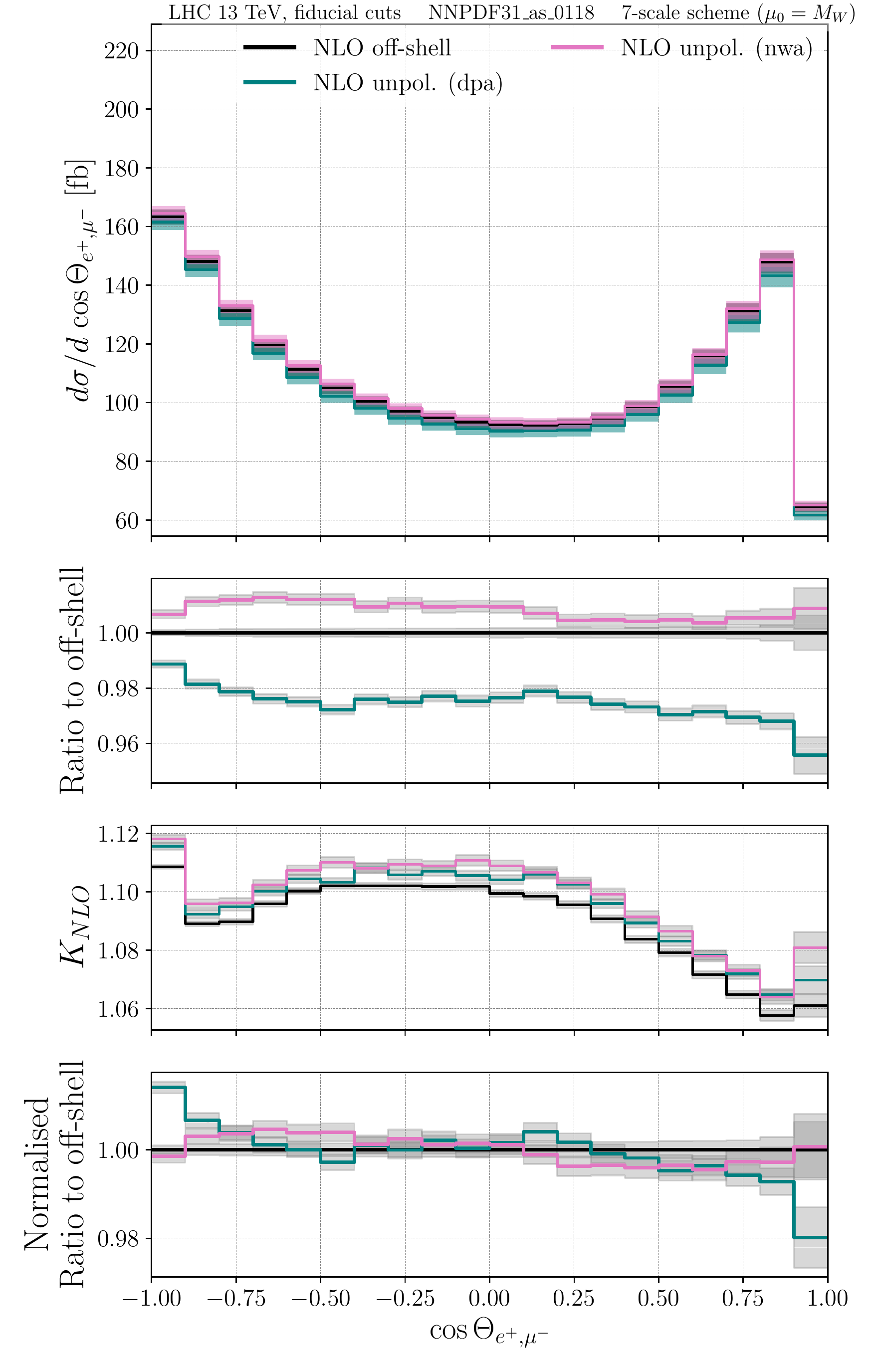}
        \label{fig:setups_cosThetaSeparation}}
  \caption{
      Comparison between off-shell calculation, DPA, and NWA for selected
      distributions. Same plot substructure as in
      \reffi{fig:setups_similarities}.
        \label{fig:setups_differences1}}
\end{figure}

As we discussed in \refse{subsec:polarised_weak_bosons}, double-pole and narrow-width
approximations consider the same set of double-resonant diagrams, however
they are different in how they generate the phase space. DPA is able to
incorporate off-shell effects via generating off-shell kinematics and subsequently
projecting it on-shell to ensure gauge invariance of the amplitude. NWA is thus
considered to be a less precise approach \cite{Denner:2020bcz}. It is therefore
instructive to compare NWA and DPA in the diboson production setting to inspect
differences in their performance.

The approximations differ on the integrated level which we discussed in
\refse{subsec:fiducial_cross_sections}.
Moving on to the differential distributions, we note that
by construction, NWA approach, is unable to describe the weak boson invariant
masses, so we will not discuss this observable. In fact, due to the absence of
single-resonant diagrams, DPA also does not describe the full off-shell
amplitude, and produces a rather symmetrical shape that is different from the off-shell
result \cite{Denner:2020bcz}.

There are distributions where NWA and DPA show the same shape but feature an
overall shift that we see on the integrated level. The symmetrised rapidity and
azimuthal angle of emission are such examples and are shown at NLO for
positron in \reffi{fig:setups_similarities}. The reader can also find a
discussion on the definition of azimuthal angle of emission in
\refapp{sec:appendix_azimuth}.

For these distributions, it makes sense
to compare them normalised by their integrated values. The bottom panes of
\reffi{fig:setups_similarities} show that the normalised shapes between DPA and
NWA agree well within their Monte-Carlo errors at NLO. This behaviour is
replicated at NNLO also with the inclusion of the loop-induced channel.

Another distribution that showcases similarities in DPA and NWA performance is
the cosine of angle between the charged leptons featured on
\reffi{fig:setups_cosThetaSeparation}. Here the approximations agree with each
other across the entire range except for the first and the last bins as one can
see from the "normalised to off-shell" plot. Fiducial
cuts affect the last bin and are responsible for its large MC errors, but in the
case of the first one there is a true difference. This is independent of QCD order and
can be observed already at LO. Perhaps, the DPA mapping underperformes in the
point where leptons are emitted in the opposite directions.

In \reffi{fig:setups_cosThetaEmission} we show a comparison between the charged
lepton emission angle in the DPA and NWA frameworks. DPA features a distribution
that is slightly further from the off-shell result. Near the first bin the
distribution is affected by the fiducial cuts and so the approximations become
further away from the off-shell calculation. The same conclusions can be reached
for angular distributions of the muon and are replicated at NNLO QCD including the
loop-induced channel.

\begin{figure}[htb]
  \centering
  \subfigure[Azimuthal separation between charged leptons (NLO)]
    {\includegraphics[width=0.48\textwidth]
      {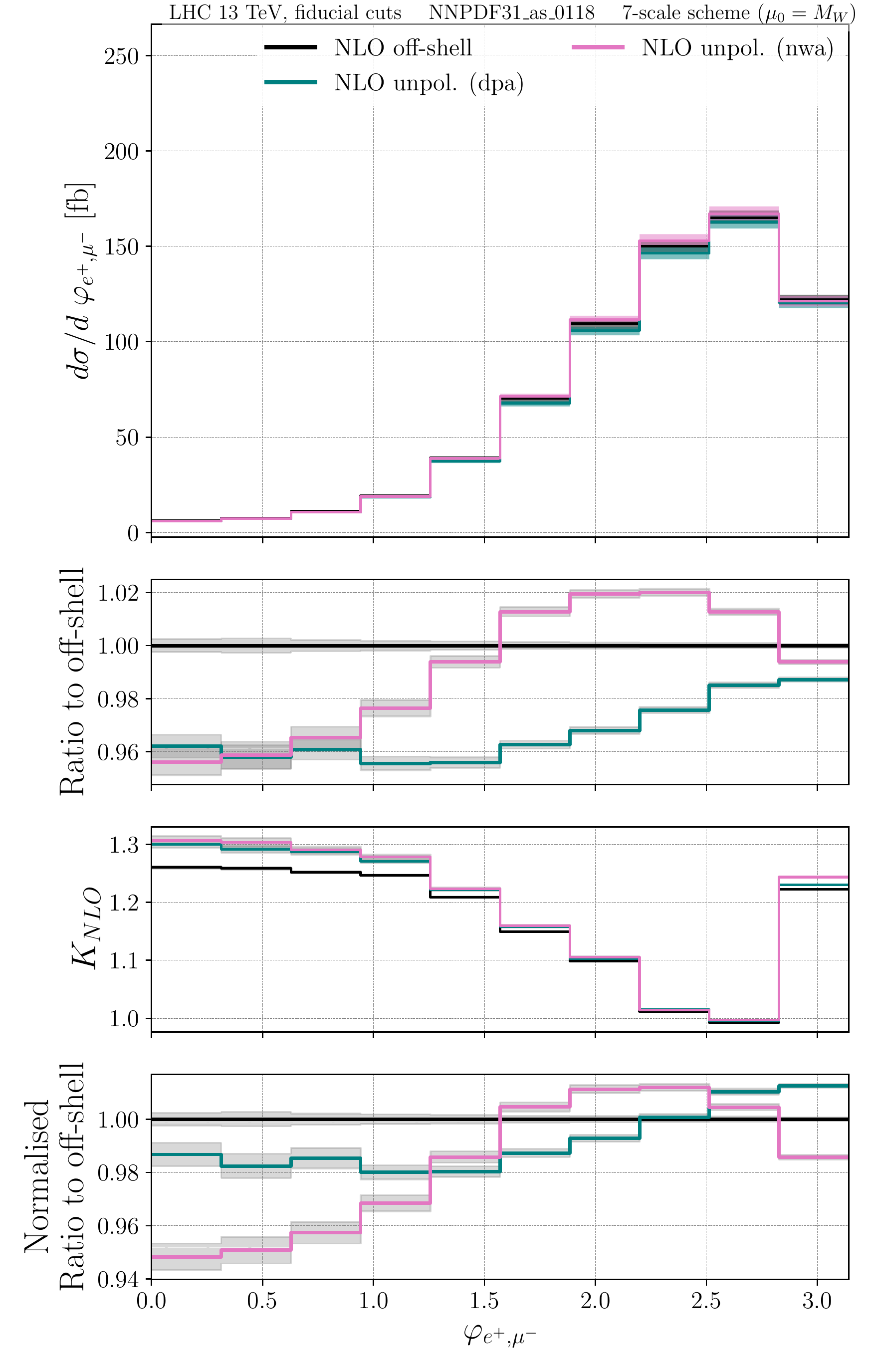}
      \label{fig:setups_differences_azimuthalSeparation}}
  \subfigure[Leading lepton transverse momentum (NLO)]
    {\includegraphics[width=0.48\textwidth]
      {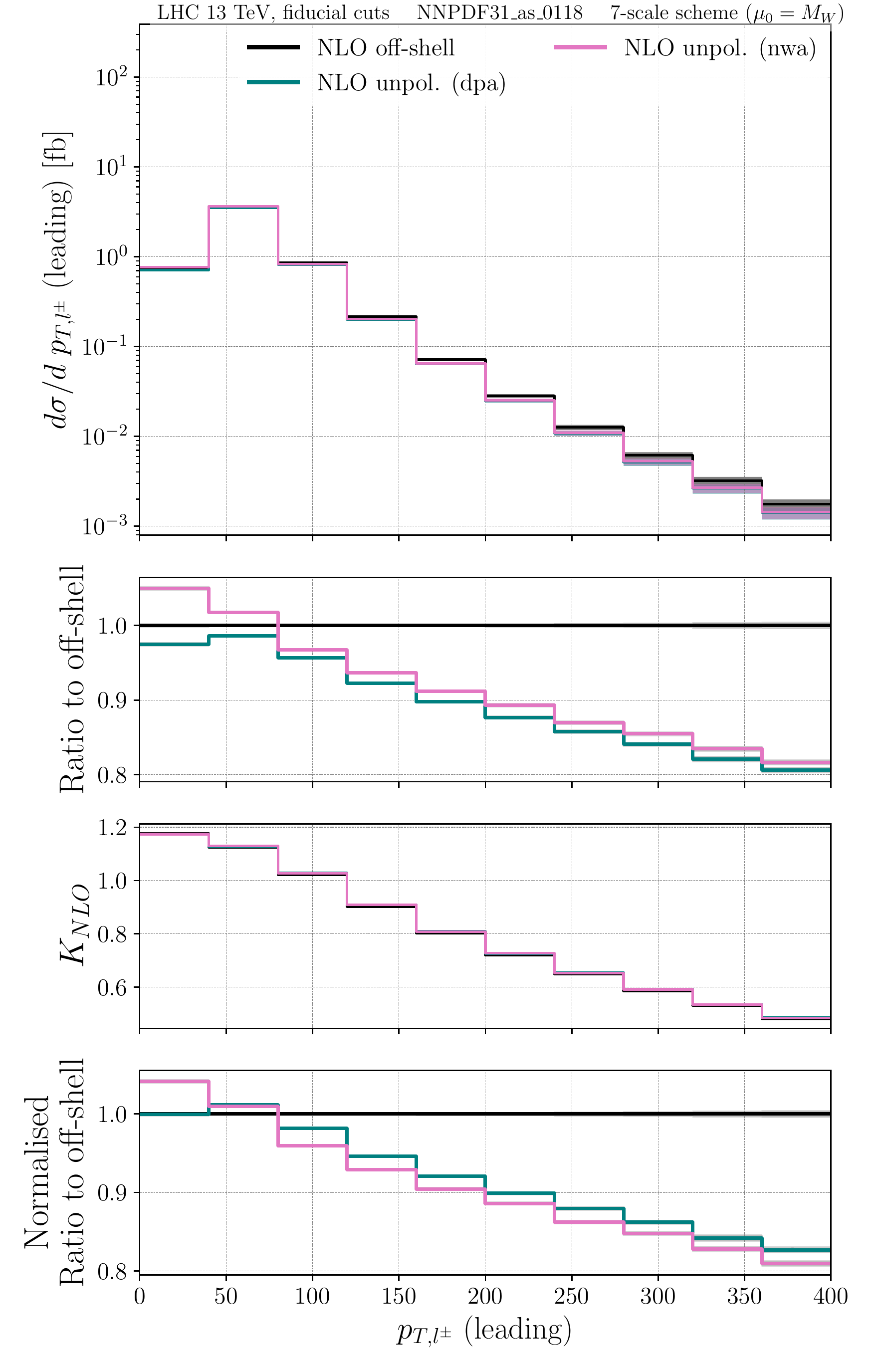}
      \label{fig:setups_differences_leptonLeading_pT}}
  \caption{
      Comparison between off-shell calculation, DPA, and NWA for selected
      distributions. Same plot substructure as in
      \reffi{fig:setups_similarities}.
        \label{fig:setups_differences}}
\end{figure}

A notable deviation can be observed between the setups at the beginning of various
transverse momentum and invariant mass distributions, which represent the bulk
of the cross section. In \reffi{fig:setups_differences_leptonLeading_pT} we present
the leading lepton $p_T$ which shows that at $p_T < 50 \GeV$ NWA overshoots the
off-shell calculation, whereas DPA undershoots. This region is the origin of the
total cross section results, as at higher $p_T$ the distribution falls nearly
exponentially. DPA and NWA differ by 2\% in the tail of the distribution and
both deviate significantly from the full off-shell calculation due to
single-resonant effects \cite{Biedermann:2016guo}. Similar effects of the two
approximations can be observed in the diboson invariant mass distribution.
$\PW$-boson transverse momentum distribution features the same interplay between
DPA and NWA, however here the approximations are closer to the off-shell result.
NNLO K-factor shapes are the same across three setups, however the loop-induced channel
appears to make a difference in the distribution tail where DPA and NWA receive 20\% larger
corrections. This effect also observed in the $\PW$-boson transverse momentum
profile.

Finally, quite an interesting difference in the behaviour between DPA and NWA can be
observed in the azimuthal angular separation between charged leptons.  In
\reffi{fig:setups_differences_azimuthalSeparation} one can see that the ratio to
full off-shell calculation is distinctly different between NWA and DPA for
$\phi_{\Pe^+,\mu^-}>0.75$. They have matching values up until this threshold and
diverge until the end of the distribution at $\phi = \pi$. In this region of
disagreement, NWA overshoots the off-shell computation. Expectedly, it appears
to be the peak of the distribution, thus contributing to the overall large NWA
integrated cross section value. This behaviour persists with introduction of
higher orders. K-factor both for NNLO corrections and for the inclusion of
loop-induced contribution has the same shape across setups and thus does not
introduce any differences to the shapes of the ratio plots.

\section{Conclusion}\label{sec:conclusion}

In this paper we compute, for the first time, polarised diboson production through NNLO
in QCD within the framework of double-pole approximation. In the calculations we considered a fiducial
phase space that emulates experimental setting which was already used in recent
theoretical and experimental studies of the polarised processes.

NNLO corrections effects are twofold. With the exclusion of the loop-induced
channel, the corrections show a controlled and predictable behaviour,
particularly in the regions that represent the bulk of the cross section. Among
notable effects we would point out significant corrections to the tail of
transverse momentum and invariant mass distributions, especially in the case of
longitudinal setups.
The scale uncertainty is brought down by a factor of 3 across all polarisation
setups.

However, the loop-induced channel changes the picture at NNLO significantly.
Being technically a leading order contribution, it massively increases the
scale uncertainty of both integrated and differential results. This
behaviour prompts for introduction of $\mathcal{O}(\alpha_s^3)$ corrections
(NLO) to the loop-induced channel, which is left for future work.

Finally, we compared the narrow-width and double-pole approximations in the
unpolarised setup. NWA overshoots the off-shell result by 1\%, while DPA is
lower by 2.5\%, which falls within their expected approximation error.
Distributions look similar in the case of rapidity and charged
lepton emission angles. We observed a significant deviation between approaches
at low transverse momenta and pointed out a local difference between the charged
lepton azimuthal separation, where NWA features a more volatile behaviour in
comparison with DPA, undershooting and overshooting the off-shell result.
Generally, we observe similar results between the methods with only slight
variation in particular observables and generally the same behaviour with
respect to their ability to describe the full off-shell result.


\section*{Acknowledgements}
We are grateful to Jean-Nicolas Lang, Giovanni Pelliccioli, and Ansgar Denner
for providing their polarisation-capable private version of \Recola and giving
us explanations. We would like to thank Jonas Lindert for technical support
with \OpenLoops. We acknowledge helpful discussions with Alexander Mitov and
useful comments from Heribertus Bayu Hartanto and the Cambridge Pheno Group.
This research has received funding from the European Research Council
(ERC) under the European Union’s Horizon 2020 Research and Innovation
Programme (grant agreement no. 683211). A.P. is also supported by
the Cambridge Trust, and Trinity College Cambridge.

\appendix
\section{Azimuthal angle of emission}
\label{sec:appendix_azimuth}

In this appendix we briefly discuss ways to define charged lepton emission
angles.

There are two reference frames commonly used in the literature: the helicity
(HE) frame and Collins-Soper (CS) frame. The helicity coordinate system is
defined in \cite{Bern:2011ie}. As we observed in the literature, it is common,
however, to simplify its construction by using a fixed reference momentum. In
order to be able to compare the distributions, we follow suit.
Here is the full algorithm we use to construct the $X'Y'Z'$ helicity frame.

Denote the first proton momentum as reference,
and define $Z'$ axis by the direction of $\PW$-boson in the
Lab frame. Then proton momenta $P_1, P_2$ are boosted into the $\PW$-boson rest
frame where they become $P'_1, P'_{2}$. We build $\vec Y'$ axis in the direction of
$[\vec P'_1 \times \vec P'_2]$ vector which is a perpendicular to the plane based on
boosted proton momenta. Finally, $X'$ axis is defined such that $X'Y'Z'$
coordinate system is right-handed, \ie $\vec X' \sim [\vec Y' \times \vec Z']$.

Another choice is the Collins-Soper frame which originates from
\cite{Collins:1977iv}. In short, the construction goes by boosting proton
momenta into the boson rest frame where they are denoted by $P'_1, P'_2$. Then
$Z'$ axis is defined as a bisection between vectors $\{\vec P'_1, -\vec P'_2\}$. The $X'$
axis is chosen as a bisection between $\{-\vec P'_1, -\vec P'_2\}$. Finally $Y'$ axis is
uniquely defined to complete the right-handed system.

We find useful the discussion of these frames in \cite{Baglio:2018rcu}, where
the authors also present the distributions corresponding to different frame
choices.

\bibliographystyle{JHEPmod}
\bibliography{polvv}

\end{document}